\documentclass[article,twocolumn,showpacs,preprintnumbers,amsmath,amssymb]{revtex4}
\usepackage[dvips]{color}
\usepackage{graphicx}% Include figure files
\usepackage{dcolumn}% Align table columns on decimal point
\usepackage{bm}% bold math
\usepackage{epsfig}

\newcommand{{\vp}}{{\vec p}}
\newcommand{{\vq}}{{\vec q}}

\newcommand{\beq}{\begin{equation}}
\newcommand{\eeq}[1]{\label{#1} \end{equation}}

\newcommand{\lton}{\mathrel{\lower.9ex
\hbox{$\stackrel{\displaystyle <}{\sim}$}}}
\newcommand{\gton}{\mathrel{\lower.9ex
\hbox{$\stackrel{\displaystyle >}{\sim}$}}}
\newcommand{\ee}{\end{equation}}
\newcommand{\bea}{\begin{eqnarray}}
\newcommand{\eea}{\end{eqnarray}}
\newcommand{\beqar}{\begin{eqnarray}}
\newcommand{\eeqar}[1]{\label{#1}\end{eqnarray}}

\newcommand{\dslash}{\partial\!\!\!/}
\newcommand{\pslash}[1]{{#1}\!\!\!/}

\newcommand{\psip}{\psi_{+}}
\newcommand{\psim}{\psi_{-}}
\newcommand{\vpp}{v_{+}}

\newcommand{\up}{u_{+}}

\newcommand{\eps}[2]{\epsilon({#1},{#2})}
\newcommand{\epsstar}[2]{\epsilon^{*}({#1},{#2})}

\newcommand{\bfk}{{\bf{k}}}

\begin{document}

\begin{flushright}
\vskip .5cm
\end{flushright} \vspace{1cm}

%\title{ A light-front wavefunction approach to heavy quark
%fragmentation in the QGP }
%\title{In medium fragmentation and dissociation of open heavy flavor}
\title{A light-cone wavefunction approach to open heavy flavor dynamics in 
QCD matter}
%\title{Applications of many-body QCD to open heavy flavor production at 
% RHIC and the LHC}
%\title{Heavy mesons in dense QCD matter: fragmentation and dissociation on 
%the light-cone}
%\title{A light-front wavefunction approach to heavy quark fragmentation and
%dissociation}
%\title{A thermal wavefunction calculation of open heavy flavor fragmentation
%and dissociation}

\author{Rishi Sharma$^1$}%
\email{rishi@lanl.gov}

\author{Ivan Vitev$^1$}
\email{ivitev@lanl.gov}

\author{Ben-Wei Zhang$^{1,2}$}
\email{bzhang@lanl.gov}

\affiliation{ $^1$ Los Alamos National Laboratory, Theoretical
Division, Los Alamos, NM 87545, USA } %

\affiliation{ $^2$ Key Laboratory of Quark $\&$ Lepton Physics (Huazhong Normal
University), Ministry of Education, China } %

\vspace*{1cm}

\begin{abstract}

We calculate the lowest order charm and beauty parton distribution functions in and
fragmentation functions into $D$ and $B$  mesons using the operator
definitions of factorized perturbative QCD. In the vacuum, we find 
the leading corrections that arise from the structure of the final-state 
hadrons. Quark-antiquark potentials extracted from the lattice are employed 
to  demonstrate the existence of open heavy flavor bound-state solutions 
in the QGP in the vicinity of the critical temperature. We provide first 
results for the in-medium  modification of the heavy quark distribution 
and decay probabilities in a co-moving plasma.
In an improved perturbative QCD description of heavy flavor dynamics in the
thermal medium, we combine $D$ and $B$ meson formation and dissociation  with
parton-level charm and beauty quark quenching to obtain predictions for the
heavy meson and non-photonic electron suppression in Cu+Cu and Pb+Pb 
collisions at RHIC and the LHC, respectively. 

\end{abstract}


\maketitle

%%%%%%%%%%%%%%%%%%%%%%%%%%%%%%%%%%%%%%%%%%%%%%%%%%%%%%%%%%%%%%%%%%%%%%%%%%

\section{Introduction}

The early production of heavy quarks makes them some of the most important
probes of the quark-gluon plasma (QGP) created in ultra-relativistic collisions of
heavy nuclei~\cite{Frawley:2008kk}. Precise and direct measurements of the multiplicities
and  differential distributions of $D$ and $B$ hadrons will soon become
available with the vertex detector upgrades at the Relativistic Heavy Ion Collider (RHIC)
and at the Large Hadron Collider (LHC). Such experimental advances will allow
to quantitatively address the key observable in high temperature Quantum Chromo-Dynamics (QCD) -
the apparent modification of particle production by energetic partons traversing a region
of hot  nuclear matter~\cite{d'Enterria:2009am,David:2009xr,Gyulassy:2003mc}
- for heavy quark jets. In the framework of perturbative QCD (pQCD),
studies in this direction have focused so far exclusively, with varying degree
of sophistication, on the loss of the leading parton or
particle energy via radiative and collisional processes~\cite{Zhang:2003wk,Wicks:2007am} 
(commonly known as jet quenching). This energy loss leads to a suppression of the observed
number of energetic particles in heavy ion (A+B) collisions relative to their number in
proton-proton (p+p, or N+N) reactions, scaled by the number of binary collisions.

It is well known that  $c$ and $b$ quark  partonic level energy loss  in
the QGP is not sufficient to explain the large suppression of non-photonic 
$e^+ + e^-$ measured at RHIC~\cite{Wicks:2007am}. These non-photonic electrons 
predominantly arise from the semi-leptonic decays of $D$ and $B$ mesons that are usually 
considered to be formed by heavy quark fragmentation outside of the medium. However, 
due to their small formation times, it is possible that $D$ and $B$ mesons could form inside 
the medium and subsequently dissociate~\cite{Adil:2006ra}. For this reason,
effective energy loss via meson dissociation in the 
QGP~\cite{Adil:2006ra,Dominguez:2008be,Dominguez:2008aa}
has also been studied in recent publications. Heavy meson dissociation approaches
have been quite successful phenomenologically~\cite{Adil:2006ra}, but the question 
of the possible thermal modifications to $D$ and $B$ hadrons has remained open.
Even though such modifications will affect fragmentation only if the time scale  
for their onset is smaller than the inverse virtuality in the parton decay,
it is surprising that to date there has been no theoretical evaluation of the
quark distribution (PDF) and fragmentation (FF) functions at $T\neq 0$. 

Heavy flavor provides the ideal testing ground for our first quantitative study
of such many-body QCD  effects because the lowest-lying Fock state in the light-cone
wavefunction expansion for $D$ and $B$ mesons~\cite{Burkardt:1995ct,Brodsky:1997de} can 
be related to the hadron rest-frame solution of the corresponding Dirac
equation~\cite{Avila:1994vi,Avila:1999aj}. Furthermore, potential model calculations can be
generalized up to temperatures of order few $T_c$~\cite{Mocsy:2007yj} using lattice QCD
input~\cite{Kaczmarek:2005ui}. In this paper we provide the first calculation of the thermal
modification of PDFs and FFs for open heavy flavor. 
As we will demonstrate below, even in equilibrium with a co-moving plasma
heavy-light quark  bound-state  solutions of the Dirac equation persist to 
temperatures $\sim 1.6 T_c$. This provides strong motivation to take the
existing studies of open heavy flavor dynamics in dense QCD matter a step 
further by combining the partonic and hadronic aspects of the observed 
high-$p_T$ cross section attenuation. 

For the realistic case of out-of-equilibrium  fast jet propagation in the thermal medium we
combine the partonic level quenching with $D$ and $B$ meson dissociation in the medium to
compute the spectrum of $D$ and $B$ mesons prior to their weak decay. We compare the relative
importance of charm and beauty quark energy loss versus meson dissociation as a function of transverse
momentum. We also show results for the non-photonic electron suppression in central Au+Au and Cu+Cu
reactions at $\sqrt{s_{NN}}=200$~GeV and central Pb+Pb collisions at   $\sqrt{s_{NN}}=5.5$~TeV.  We
discuss the prospects  of testing the pQCD theory by  upcoming open heavy flavor measurements
at  RHIC and the LHC.

Our paper is organized as follows. In Section~\ref{baseline} we present a
detailed baseline calculation of charm and beauty quark production, inclusive of
the known cold nuclear matter effects, and expose the limitations of the purely
partonic approach to their final-state dynamics in the QGP.  In
Section~\ref{hqfrag} we express the parton distribution and decay probabilities of
heavy quarks in hadrons in terms of the light-cone wavefunctions and calculate
the $c$ and $b$ quark FFs.  In Section~\ref{potmod}  we evaluate  the heavy
meson wavefunctions in the vacuum and in the vicinity of the QCD phase
transition and present numerical results for the corresponding distribution and
decay probabilities.  A complete calculation that includes $D$ and $B$ meson
dissociation and $c$  and $b$ quark quenching, relevant to current and future
heavy flavor/non-photonic electron measurements, is given in
Section~\ref{application}. A summary and conclusions are presented in
Section~\ref{conclude}.  Background information and some of the technical aspects 
of this work are given in the Appendix. 
Appendix~\ref{Quantize} describes notation and quantization of parton fields.
In Appendix~\ref{Instant} we elucidate the relation between the instant and 
light-cone forms of a heavy meson wavefunction. Demonstration of the smallness
of the thermal light quark pick-up correction to the heavy quark FFs is given
in Appendix~\ref{pickup}. 
Appendix ~\ref{DiffEq} presents an analytic solution for the evolution of the heavy 
quark and meson system  for time-averaged fragmentation and dissociation
rates. 

\section{A timeless limit }
\label{baseline}

To correctly assess the discrepancy between the expected heavy quark
quenching and the measured non-photonic electron
suppression~\cite{Adare:2006nq,Abelev:2006db} improved treatment of
cold nuclear matter effects is necessary. A previous study of $D$ and $B$ meson 
production in p+A reactions~\cite{Vitev:2006bi} indicated that these will 
affect the nuclear modification ratio ($R_{AB}$) for energetic particles~\cite{Gyulassy:2003mc}. 
This ratio is defined as follows:
\begin{eqnarray}
\label{RAB}
&& \!\!\!\!\!\!\! R_{AB}({\bf {p}})  = \frac{ d\sigma_{AB} }{ dy  d^2{\bf p}  } 
{\bigg /}   \left(N_{\rm bin.}^{AB}\right)  \, 
\frac{ d\sigma_{NN} }{ dy  d^2{\bf p}  }    \; ,
\end{eqnarray}  
and is used as one of the main observables to characterize the properties of the system created in
heavy ion reactions. In Eq.~(\ref{RAB}), $\sigma_{AB}$ refers to the production cross-section in
collisions between two heavy ions A and B, $\sigma_{NN}$ refers to the cross-section in
collisions between two nucleons N, and $N_{\rm bin.}^{AB}$ refers to the number of binary N+N
collisions in an A+B reaction.  $N_{\rm bin.}^{AB}$ is determined by using a 
Glauber model~\cite{Gribov:1968jf}. It is, therefore, 
important  to establish the role of shadowing, Cronin effect and cold nuclear matter energy loss 
in the analysis of open heavy flavor suppression. Note that in our notation 
${\bf p}\,, \; {\bf k}\, \cdots $ are 2D transverse vectors and, for example, $p_T = |{\bf p}|$.

At the partonic level, the cross sections per elementary nucleon-nucleon collision,
including the ones for charm and beauty quarks, can be calculated  as follows:
\begin{eqnarray}
\label{single}
&& \!\!\!\!\!\!\!  \left(N_{\rm bin.}^{AB}\right)^{-1}
\frac{ d\sigma^{q,g}_{AB} }{ dy  d^2{\bf p}  }   =
K  \sum_{abcd}  \int  dy_d \int  d^2 {\bf k}_{a}  d^2 {\bf k}_{b} \;
\frac{ f({k}_{Ta}) f({k}_{Tb}) }{|J({k}_{Ta},{k}_{Tb} )|}
\nonumber \\
&& \!\!\!\!\!\!\!\! \times \, \frac{\alpha_s^2(\mu_r)}{2S} 
|\overline {M}_{ab\rightarrow cd}|^2
\frac{\phi_{a/N} \left(\frac{ {\tilde{x}}_a}{1-\epsilon_a},\mu_f\right)
\phi_{b/N}\left(\frac{{\tilde{x}}_b}{1-\epsilon_b},\mu_f\right) }
{ {\tilde{x}}_a {\tilde{x}}_b } \,
\;. \qquad
\label{LO}
\end{eqnarray}
In Eq.~(\ref{LO}) $\bfk_a,\;\bfk_b$ refer to the transverse momenta of the two partons $a,\;
b$, respectively, that participate in the hard collision and $x_a,\; x_b$ is the fraction of the light cone 
momentum of the nucleon N carried by these. $ {M}_{ab\rightarrow cd}$ denotes the
matrix element of the hard collision of $a$ and $b$ to give rise to partons $c$ and $d$. $\phi(x,\mu)$
refers to the parton distribution functions that give the probability to find a parton with light cone
momentum fraction $x$ of the nucleon light cone momentum.  We set
the factorization and renormalization scales to $\mu_f = \mu_r = m_T = (m_{q,g}^2+{{p_T}}^2)^{1/2}$
and the Jacobian $J$ reads:
\beq
J_{x_a,x_b} ({k}_{Ta},{k}_{Tb} ) =
\frac{S}{2} \left(1-\frac{k_{Ta}^2k_{Tb}^2}{x_a^2S x_b^2S} \right) \; ,
\eeq{jac}
subject to the hard scattering constraint $k_{Ta,b} < x_{a,b} \sqrt{S}$.
We have generalized the pQCD collinear factorization approach to heavy flavor 
production~\cite{Vitev:2006bi,Olness:1997yc} to account for the
non-zero parton transverse momentum distribution $f({k}_{Ta,b})$ in the hadron
wavefunction with $\langle {k}_{Ta,b}^2 \rangle_{NN} = 0.9$~GeV$^2$~\cite{Owens:1986mp}. 
The phenomenological next-to-leading order $K$ factor in Eq.~(\ref{LO})
cancels in the observable  $R_{AB}({{p_T}})$, Eq.~(\ref{RAB}). Finally, we note that the momentum fraction $x$ 
that appears in Eq.~(\ref{single}) is modified due to many-body QCD scattering effects that we
discuss below.

{\it Cronin effect.} In p+A and A+A reactions the Cronin effect can be taken into account 
in the ${\bf{p}}$-differential cross sections by including the $\bfk$ broadening of
incoming partons that arises from initial-state 
scattering~\cite{Accardi:2002ik,Vitev:2003xu}.
This transverse momentum broadening in heavy ion collisions is simplest to evaluate 
if $f({k}_{Ta,b})$ is of a normalized Gaussian form due to the additive dispersion property: 
The width of the transverse distribution
function $f$ grows relative to its value in N+N collisions as follows:   
\beqar   
&& \langle {k}_{Ta,b}^2 \rangle_{AB} = \langle {k}_{Ta,b}^2 \rangle_{NN}
+  \langle {k}_{Ta,b}^2 \rangle_{IS}\, , \; \nonumber \\ 
&& \langle {k}_{Ta,b}^2 \rangle_{IS} = \left\langle \frac{2 \mu^2 L}{\lambda_{q,g}}  
\right\rangle \xi \; .
\eeqar{Cron} 
Specifically, in Eq.~(\ref{Cron})  $\mu^2 = 0.12$~GeV$^2$, 
$\lambda_g = (C_F/C_A) \lambda_q = 1$~fm 
and  $\xi \sim $few is a numerical factor that accounts for the enhancement of the 
broadening coming from the  power-law tails of the  Moliere multiple scattering~\cite{Accardi:2002ik,
Vitev:2003xu}. 

{\it Cold nuclear matter-energy loss.}   Initial-state  multiple
collisions also give rise to  radiative energy loss that is sensitive to the
quark or gluon   mass. Theoretical derivation of the stopping  power of 
large nuclei for 
partons was given in Ref.~\cite{Vitev:2007ve} and numerical calculations 
were carried out for the same squared transverse momentum 
transfers and parton mean free paths given above.  
If fast quarks or gluons lose a fraction 
$\epsilon_{a,b}= \frac{\Delta E^{\rm{rad}}_{a,b}}{ E_{a,b} }$ 
of their energy prior to the
hard scattering, to satisfy the same final-state kinematics they must initially 
carry a larger fraction of the colliding hadron
momentum and, correspondingly, a larger value of $x$.   This can be implemented in 
Eq.~(\ref{LO}) as follows:
\beq {\tilde{x}}_{a} \rightarrow \frac{ {\tilde{x}}_{a}}{1-\epsilon_{a}} \;, 
\quad   {\tilde{x}}_{b} \rightarrow \frac{ {\tilde{x}}_{b}}{1-\epsilon_{b}}\;, \quad  
{\tilde{x}}_{a,b} \leq 1 \; , 
\eeq{eloss}
in the parton distribution functions  $\phi_{a,b/N}({\tilde{x}}_{a,b}, \mu_f)$. 
$\tilde{x}_{a,\;b}$ will be defined in terms of $x_{a\;, b}$ in a moment when we 
consider dynamical shadowing. We note that the theoretical calculation of cold nuclear matter energy loss 
is most easily performed in the target nucleus rest frame where the incident 
partons are typically of very high energy~\cite{Vitev:2007ve}.

{\it Dynamical shadowing.}   Last but not least, power-suppressed~\cite{Qiu:2001hj}
resummed~\cite{Qiu:2003vd,Qiu:2004da} coherent final-state scattering of the
struck partons that leads to shadowing  in the observed cross sections is included
via~\cite{Vitev:2006bi}:
\beqar
\label{shada}
&&   \tilde{x}_{a} = x_a\left(1+\frac{\xi_d^2(A^{1/3}-1)}{-\hat{t}+m_d^2}\right)\; , 
\\[.5ex]
&&  \tilde{x}_{b} = x_b\left(1+\frac{\xi_c^2(A^{1/3}-1)}{-\hat{u}+m_c^2}\right)\; . 
\label{shadb}
\eeqar{shad}
In Eqs.~(\ref{shada}) and (\ref{shadb})  $\hat{t}, \; \hat{u}$ are the relevant  
Mandelstam variables at the partonic level and $m_c$, $m_d$ are the masses of the struck target
partons. $A$ is the atomic number of the nucleus. A consistent picture of cold nuclear matter
effects requires $ (\xi^2)_{q,g} A^{1/3} \approx ( 2 \mu^2 L/ \lambda )_{q,g}$ in minimum bias
reactions, which yields $(\xi^2)_q \approx 0.12$~GeV$^2$ and $(\xi^2)_g \approx 0.27$~GeV$^2$.

Our results, reported here, bring the treatment of cold nuclear matter effects
for open heavy flavor production on par with their implementation in the study
of light hadron and direct photon final states~\cite{Vitev:2008vk}. 
Linear dependence of cold nuclear matter effects on the mean nuclear thickness
of the interaction region is implicit in our approach and reflects the actual 
theoretical findings.

We now turn to the final-state quark and gluon dynamics that is governed by the
formation times of the plasma~\cite{Dumitru:2006pz}, $\tau_0$, and of the 
hadrons, $\tau_{\rm{form}}$~\cite{Adil:2006ra}. Using the uncertainty principle, 
one can evaluate:  
\begin{eqnarray}
\tau_{\rm{form}}  &\simeq &  \frac{2z(1-z) E_T }{ {\bf k}^2
+ (1-z)m_h^2 - z(1-z)m_q^2   } \;, \quad
\label{tfrag}
\end{eqnarray}
which grows with underlying jet energy $(E_T)$, depends on the fractional momentum 
$z = E_h / E_{q,g}$ carried by the leading hadron and is inversely proportional
to the square of its mass $(m_h)$. We take the formation time of the QGP to be 
$\tau_0 = 0.6$~fm and $\tau_0 = 0.3$ fm at RHIC and LHC, respectively, compatible
with hydrodynamic studies~\cite{Hirano:2008hy}. Simulations of 
the early stages of the reaction~\cite{Dumitru:2006pz} suggest that the 
transverse energy density of the partonic  system  builds through  a short 
exponential growth. Therefore, for $\tau < \tau_0$ particles 
will be largely unaffected by the medium.

For $\tau \geq \tau_0$  the properties of the soft gluon-dominated matter, 
created in heavy ion reactions with multiplicity proportional to the 
participant density of the colliding nuclei:  
\beq
\frac{dN^g}{dy d^2 {\bf x}} ({\bf x})  = 
\kappa \frac{dN^{part.}}{ d^2 {\bf x}} ({\bf x}) \; ,    
\eeq{partdens}
can be constrained by the experimentally measured charged particle pseudorapidity 
density~\cite{Back:2001ae}.  Integrating over the position in the transverse 
plane ${\bf x}$ in Eq.~(\ref{partdens}) we can determine $\kappa$ from 
simple parton-hadron duality:
\beq
\frac{dN^g}{dy} = \frac{3}{2} \left| \frac{d \eta}{dy} \right 
| \frac{dN^{ch}}{d\eta} \; . 
\eeq{dual}
Here, $dN^g/dy$ refers to the gluon rapidity density, and $\eta = -\log(\tan(\theta/2))$ is the
pseudorapidity, where $\theta$ is the angle of the observed particle with respect to the beam axis.
For example, in central Cu+Cu and Au+Au collisions 
at RHIC, at the average nucleon-nucleon collision energy, $\sqrt{s_{NN}}=200$~GeV,
$\frac{dN^g}{dy} = 250  $ and 1050, respectively, and in central 
Pb+Pb collisions at the LHC at $\sqrt{s_{NN}}=5.5$~TeV one expects
$\frac{dN^g}{dy} \simeq 2800  $.  The local time- and position-dependent 
temperature  for an isentropic Bjorken expansion can be readily obtained:
\begin{equation}
T(t, {\bf x}) = \  ^3\!\sqrt{ {\frac{\pi^2}{16 \zeta(3)}  
\frac{1}{t} \frac{dN^g}{dy d^2 {\bf x}} ({\bf x}) } }  \; .
\label{tempdet}
\end{equation}
In Eq.~(\ref{tempdet}) we have replaced the proper time $\tau(t,x^3)=\sqrt{t^2-(x^3)^2}$, where
$x^3$ is the position coordinate along the axis of the collision, by the time $t$. This is a
good approximation for $x^3$ small compared to $t$, meaning rapidities $y$ small
compared to $1$. The Debye screening scale, $m_D = g T$,  and the relevant
gluon mean free path  $\lambda_g = 1/ \sigma^{gg} \rho$
with $\sigma^{gg} = (9/2) \pi \alpha_s^2 / m_D^2$, that enter 
jet quenching calculations can then be evaluated.

\begin{figure}[t]
\begin{center}
%\vspace*{-1.in}
%  \psfig{file=HeavyQuenchNew.eps,height=3.8in,width=3.2in,angle=0}
\includegraphics[width=3.2in,angle=0]{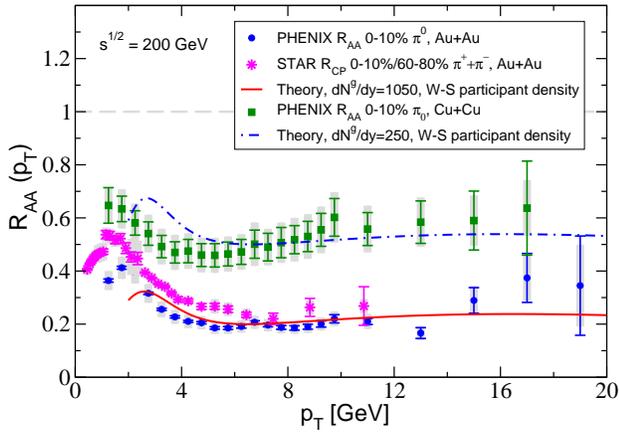} 
%\vspace*{-1.in}
\caption{ Perturbative QCD calculation of the  $\pi^0$ and
$\pi^+ + \pi^-$ quenching in central Au+Au 
and  Cu+Cu  collisions at RHIC at $\sqrt{s_{NN}}=200$~GeV. 
Pion suppression data from PHENIX~\cite{:2008cx,Adare:2008cg} and
STAR~\cite{Abelev:2007ra} is also shown.}
\label{success}
\end{center}
\end{figure}

In contrast to the soft gluons discussed above, large $Q^2$ parton processes follow
a binary collision density distribution, $\propto \frac{dN^{bin.}}{d^2 {\bf{x}}}$,
that is peaked near the center of the collision geometry. In Ref.~\cite{Vitev:2008rz}
a complete and numerically intensive calculation of jet production and 
propagation through the plasma  was carried out to confirm that the
energy loss pattern is determined by the jets originating near the peak of the
binary collision density. Therefore, we here focus only on these hard partons, 
use a running coupling constant at the radiation vertex and describe the strength 
of the interactions of the jet with the medium by an effective value 
$\alpha_s = 0.28-0.32$. Any uncertainty associated with higher orders
in opacity and peripheral jets will lead to somewhat larger values of $\alpha_s$.
The energy loss calculation itself is performed in the framework of the GLV
approach~\cite{Gyulassy:2003mc,Vitev:2007ve} and the 
Landau-Pomeranchuk-Migdal destructive interference effect in QCD is
fully accounted for. We evaluate the probability distribution $P(\epsilon)$ 
for  the parton to  lose a fraction of its energy loss 
$\epsilon = \sum_i \frac{\Delta E_i}{E}$ due to multiple
gluon emission. The quenched quark or gluon spectrum is 
then readily obtained:
\begin{eqnarray}
&&
\!\!\!\!\frac{ d\sigma^{q,g\; Quench}_{AB} ({\bf p})}{ dy d^2{\bf p}  }   
= \int_0^1 d\epsilon \; P(\epsilon) \frac{1}{(1-\epsilon)^2}  
\frac{d\sigma^{q,g}_{AB}\left(\frac{{\bf p}}{1-\epsilon}\right) }
{ dy d^2{\bf p}  }
\;. \qquad
\label{Quench}
\end{eqnarray}
In Eq.~(\ref{Quench})  the factor $(1-\epsilon)^{-2}$ arises from 
the transverse phase space Jacobian $|d^2{\bf p}/d^2{\bf p}^{Quench}  |$.
Hadronization is performed using the appropriate vacuum fragmentation
function $D_{h/q,g}(z,\mu_{fr})$ as follows: 
\begin{eqnarray}
\frac{ d\sigma^{h}_{AB} ({\bf p})}{ dy d^2{\bf p}  }  &=& \sum_{q,g} \int_0^1 dz \;
D_{h/{q,g}}(z,\mu_{fr})  \nonumber \\
&&   \times \frac{1}{z^2}  \frac{d\sigma^{q,g\; Quench}_{AB}\left(\frac{{\bf p}}{z}\right) }
{ dy d^2{\bf p}  }
\;, \qquad
\label{Frag}
\end{eqnarray}
where $z$ refers to the fraction of the light cone energy of the initial parton $q,\;g$, carried by
the hadron $h$.

\begin{figure}[t]
\begin{center}
%\vspace*{-1.in}
%  \psfig{file=HeavyQuenchNew.eps,height=3.8in,width=3.2in,angle=0}
\includegraphics[width=3.2in,height=3.8in,angle=0]{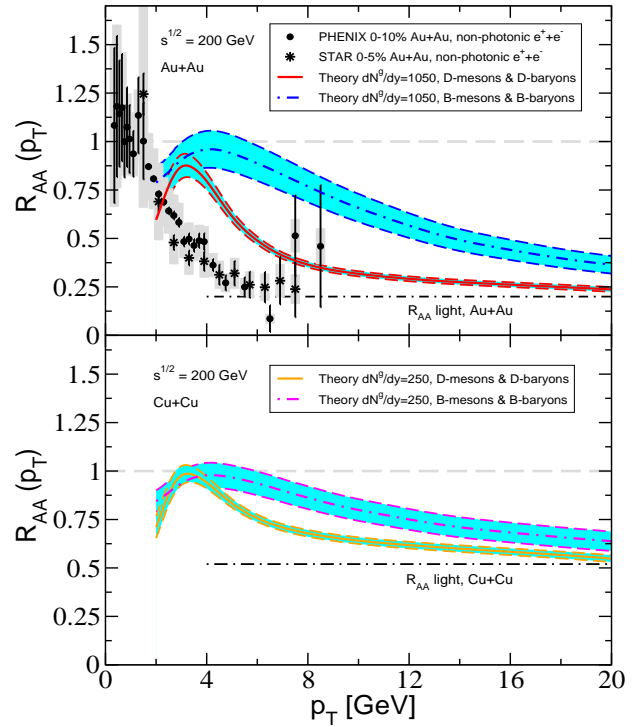} 
%\vspace*{-1.in}
\caption{ Perturbative QCD calculation of the suppressed 
$D$ and $B$ meson production cross sections in central Au+Au (top panel)
and  Cu+Cu (bottom panel) collisions is compared to the magnitude of
light pion  quenching at $\sqrt{s_{NN}}=200$~GeV. The effect of parton
mass on the cold nuclear matter effects is illustrated via uncertainty
bands. Non-photonic electron data from PHENIX~\cite{Adare:2006nq} and
STAR~\cite{Abelev:2006db} is also shown.}
\label{problem}
\end{center}
\end{figure}

Eqs.~(\ref{LO}), (\ref{Quench})  and (\ref{Frag}) summarize the 
traditional approach to jet quenching that has been  very
successful in the description of the observed attenuation of
light hadron production in A+A reactions. Figure~\ref{success} 
illustrates the level of agreement that can be achieved between  
the perturbative QCD theory, described in this section, and the
experimental measurements of $\pi^0$~\cite{:2008cx,Adare:2008cg} 
and $\frac{1}{2}(\pi^+ + \pi^-) $~\cite{Abelev:2007ra} at RHIC. 
In central Au+Au and Cu+Cu reactions we used gluon rapidity densities 
$dN^g/dy = 1050$ and $dN^g/dy = 250$. 
Identical calculation for open heavy flavor production in 
heavy ion collisions at $\sqrt{s_{NN}} =200$~GeV is presented in
Fig.~\ref{problem}.  It is easy to see that Cronin enhancement plays an important 
role at intermediate $p_T \sim 4$~GeV. The bands show the sensitivity of 
cold nuclear matter effects when we vary all parton masses in the range 
from $0$ to $m_c,m_b$, respectively. At mid-rapidity, $y=0$, the effect of
high-twist shadowing is small and the uncertainty band is dominated by initial state
energy loss. Cold nuclear matter effects amplify the disparity between 
the heavy and light parton quenching. We have schematically illustrated the latter
by showing the  magnitude of pion attenuation from  Fig.~\ref{success}.

In contrast, the experimental results on non-photonic
electrons~\cite{Abelev:2006db,Adare:2006nq} $R_{AA}^{e^\pm} \sim
R_{AA}^{\pi^0}, \; p_T >5$~GeV  are in clear contradiction 
with the small quenching of $B$ mesons that give an
increasingly important $\geq 50\%$ contribution to non-photonic $e^+ + e^-$ in 
this region~\cite{Adil:2006ra,Gang}. An insight into the 
possible cause for this well-known by now discrepancy can be
gained by examining Eq.~(\ref{tfrag}). The small pion mass ensures 
that the parent light quarks and gluons fragment outside of the QGP 
in accord with the traditional picture of jet quenching. The large
$D$ and $B$ meson mass, however, implies that charm and beauty
quarks will fragment inside the hot and dense medium: $ \tau_{\rm form} 
\propto 1/m_h^2$.  Consequently, the competition between heavy 
meson  dissociation and the $c$ and $b$ quark decay in the QGP
is a likely physics mechanism that may
naturally lead to attenuation of the beauty cross section as 
large as that for charm~\cite{Adil:2006ra}.

%%%%%%%%%%%%%%%%%%%%%%%%%%%%%%%%%%%%%%%%%%%%%%%%%%%%%%%%%%%%%%%%%%%%%%%%%%%%

%%%%%%%%%%%%%%%%%%%%%%%%%%%%%%%%%%%%%%%%%%%%%%%%%%%%%%%%%%%%%%%%%%%%%%%%%%%%

\section{Distribution and fragmentation functions of heavy quarks}
\label{hqfrag}

Simulations of heavy quark fragmentation and dissociation in the QGP require
knowledge of the corresponding parton decay and distribution probabilities.
These are related to the wavefunctions of their parent or decay hadrons  that can 
be represented as follows:  
\begin{eqnarray}
&& |\vec{P}^+;J \rangle  =   {a}_h^\dagger(\vec{P}^+;J)  |0 \rangle
= \sum_{n = 2(3)}^\infty  \int \prod_{i=1}^n  \frac{d^2{\bf k}_i}{\sqrt{(2\pi)^{3}}}
\frac{dx_i}
{ \sqrt{2 x_i}}  \nonumber \\ 
&& \times \, \psi(x_i,{\bf k}_i ; \alpha_i) \,  \delta \left(\sum_{j=1}^n x_j -1 \right) \, 
\delta^2  \left( \sum_{j=1}^n{{\bf k}_j} \right)  \nonumber \\
&&  \times |\cdots  {a}^\dagger_{q_i}(x_{q_i}\vec{P}^+ + {\bf k}_{q_i}, \alpha_{q_i} ) \cdots 
 {b}^\dagger_{\bar{q}_j}(x_{\bar{q}_j}\vec{P}^+ + {\bf k}_{\bar{q}_j}, \alpha_{{\bar{q}_j} })
\cdots     \nonumber \\
&& \cdots   {c}^\dagger_{{g}_k}(x_{{g}_k}\vec{P}^+ + {\bf k}_{{g}_k}, \alpha_{{{g}_k} } ) 
\cdots \rangle \; .
\label{Mp}
\end{eqnarray}   
Here,  $\vec{P^+} \equiv (P^+,{\bf P})$  are the large light-cone momentum 
and transverse momentum components of the hadron, and the momenta of the partons 
are given by $(x_iP^+,x_i{\bf P}+{\bf k}_i)$. Our conventions for the 
quantization of the fermion and vector fields, including notation and 
commutation relations for the creation and annihilation operators 
$a^\dagger$, $a$, $\cdots$, are given in Appendix~\ref{Quantize}. In Eq.~(\ref{Mp})
$\alpha_{q_i}$ is a set of additional relevant quantum numbers, such as helicity 
and color. From the normalization of the meson state of fixed projection $\lambda$
of the total angular momentum $J$: 
\beq
\langle \vec{P}^+;J  |\vec{P}^{+ \prime};J \rangle
= 2P^+ (2\pi)^3 \delta(P^+-P^{+ \prime}) \delta^2 ({\bf P}-{\bf P}^\prime)
\delta_{\lambda \lambda^\prime}\;,  
\eeq{mesonnorm}
an integral constraint  on the norm of the light-cone wavefunctions 
$\psi(x_i,\bfk_i;\alpha_i)$ can be derived.

In Eq.~(\ref{Mp}) proper construction  of the non-perturbative lowest order 
Fock component in the  expansion  of a hadronic state in a quark and gluon  basis,
$q\bar{q}$  for mesons and $qqq$ for baryons, appears  to be the most 
important. When properly constructed,  higher order components that are generated 
via  parton  splitting,  will  have the correct quantum numbers through 
preservation of all relevant  global symmetries in QCD. These components become 
dominant in the small momentum fraction  $x$ region of sea quarks and gluons. 
On the other hand, for $x \sim {\cal O}(1)$  the momentum distribution of valence
quarks can be determined  from the relativistic quark model. For heavy quarks ($Q$), 
perturbative splitting is suppressed by factors ${\cal O}(\Lambda_{QCD}^2/M_Q^2 )$  
relative to light quarks ($q$) and gluons ($g$) and if we focus on the charm and beauty 
distribution and fragmentation functions the lowest order $Q\bar{q}$ (or $\bar{Q}q$)  
Fock component for heavy mesons will be a good starting point for our calculation:
\begin{eqnarray}
&& |\vec{P}^+;J \rangle  %= a_h^\dagger(\vec{P}^+;J)  |0 \rangle
= \int \frac{d^2{\bf k}}{(2\pi)^{3}} \frac{dx}{ 2\sqrt{ x(1-x)}}
\frac{M(j)_{s_1s_2}}{\sqrt{2}}  \frac{\delta_{c_1c_2}}{\sqrt{3}} \,
\psi(x,{\bf k})\nonumber \\
&& \times a_Q^{\dagger\; s_1 c_1 }(x\vec{P}^++{\bf k})  b_q^{\dagger \; s_2 c_2  }
\bigl((1-x)\vec{P}^+-{\bf k}\bigr)  |0 \rangle  \; . 
\label{Mp1}
\end{eqnarray}
In Eq.~(\ref{Mp1}) $s_1, \;s_2$ refer to the spins of $Q$ and $\bar{q}$ and we have used the notation:
\begin{eqnarray}
&& M(P)_{s_1s_2} = (\chi)_{s_1s_2} = \left (
\begin{matrix} 0 &1 \\ -1& 0  \end{matrix} \right )_{s_1 s_2 }\;,    \\
&& M(V)_{s_1s_2} = \sum_i(\epsilon_i(\lambda) \sigma_i \chi )_{s_1 s_2 } \;,
\label{Mvec}
\end{eqnarray}
to ensure the correct spin structure of  pseudoscalar $^1S_0$ and vector 
$^3S_1$ states and proper normalization ~\cite{Ma:1994zt}. 
A set of polarization vectors for $J=1$ in Eq.~(\ref{Mvec}) in the meson rest frame is given by:
\begin{eqnarray}
\vec{\epsilon}_{\pm 1} =  \mp \frac{1}{\sqrt{2}} 
\left(\begin{array}{ccc}  1  \\ \pm i \\ 0 \end{array} \right) \;, 
\vec{\epsilon}_{3} =   
\left(\begin{array}{ccc}  0 \\ 0 \\ 1 \end{array} \right) \;. 
\label{polar}
\end{eqnarray}
The color  singlet structure is ensured by $\delta_{c_1c_2}$ where $c_1,\; c_2$ are the color 
indices of $Q$ and $\bar{q}$, respectively. Following Eq.~(\ref{mesonnorm})
we can derive the normalization for each helicity $\lambda$:  
\beq
\frac{1}{2 (2\pi)^{3} } \int dx d^2{\bf k}  \;
| \psi(x,{\bf k}) |^2 = 1 \; .
\eeq{lonorm}
Note that the meson wavefunction can also be written in an instant form. 
For example, in the rest frame of the meson: 
\begin{eqnarray}
&& |\vec{{P}}=0;J \rangle_{h} 
= \int \frac{d^3\vec{{q}}}{(2\pi)^{3}}
\frac{M(j)_{s_1s_2}}{\sqrt{2}}  \frac{\delta_{c_1c_2}}{\sqrt{3}} \,
f(\vec{{q}})\nonumber \\
&& \hspace*{2cm}\times \sqrt{\frac{E_h}{2E_1E_2}}a_Q^{\dagger\; s_1 c_1 }
(\vec{{q}})  b_q^{\dagger \; s_2 c_2  } (-\vec{{q}})  |0 \rangle
\label{MpI} \; , \qquad 
\end{eqnarray}
where ${\vec{q}}$ represents the relative momentum between quarks
and $f({\vec{q}})$ is the Fourier transform of the meson wavefunction 
in position space. $E_h=\sqrt{\vec{ P}^2+m_h^2}$ is the meson energy 
and  $E_i=\sqrt{ \vec{  q}_i^2+m_i^2}$ are the quark energies. 
Conversion between the instant and light-cone forms is facilitated  
if harmonic oscillator, or Gaussian, approximation is made for the 
shape of the wavefunction $f(\vec{{q}})$~\cite{Ma:1993ht}, which 
allows to easily  ensure identical transverse width 
$\langle {\bf k}^2 \rangle$~\cite{Adil:2006ra} for both forms.

In the light-cone gauge $A^+ = 0$, the heavy quark distribution function
is given by~\cite{Collins:1981uw}:
\begin{eqnarray}
\phi_{Q/h}(x) &=&\int \frac{dy^-}{2\pi} e^{-i x P^+ y^-}
\nonumber \\
&& \times \langle P^+ |\bar{\psi}_Q(y^-,{\bf 0}) 
 \frac{\gamma^+}{2}\psi_Q(0,{\bf 0}) |  P^+ \rangle 
\label{PDFdef} \;, 
\end{eqnarray}
where $\psi_Q$ is the heavy quark field operator, $|P^+\rangle$ are 
appropriately normalized hadronic states
with large light-cone momentum $P^+$, and the average over the 
hadron spin/polarization states is implicit. 
For both the pseudoscalar and the
vector cases, substituting Eq.~(\ref{Mp1}) in the perturbative QCD definition,  
we proceed to a straightforward evaluation at tree level. We first carry out the 
contractions of the heavy and light quark creation and annihilation operators, 
perform the color sums and the integration over the internal kinematic 
variables to arrive at an intermediate result:
\beqar
\phi_{Q/h}(x)&\propto&  \int \frac{d^2{\bf k}}{(2\pi)^3}\frac{dx_Q}{2x_Q} 
\int \frac{dy^-}{2\pi} e^{-i(x - x_Q)P^+y^-} \, 
 \nonumber \\
& & \times |\psi(x_Q, {\bf k})|^2 \, \frac{1}{2J+1}\sum_{\lambda}
\bigl(M(\lambda)M^\dagger(\lambda)\bigr)_{s_1 s_2}   \nonumber \\
&& \times \frac{1}{2} \bar{u}^{s_1}(x_QP^+)\frac{\gamma^+}{2} u^{s_2}(x_QP^+)  \; , \nonumber 
\label{interemediate}
\eeqar{intermediate}
where $u^s(p)$ are Dirac spinors defined in Appendix~\ref{Quantize}. With the polarization average
yielding $\delta_{s_1 s_2}$ and the  spin sum resulting in $x_QP^+$  we readily obtain the final
result: 
\begin{eqnarray}
\phi_{Q/h}(x) = \frac{1}{2 (2\pi)^3}
\int dx_Q d^2{\bf k}  \; | \psi(x_Q,{\bf k}) |^2 \delta(x_Q-x) \; . \quad
\label{PDFdef2}
\end{eqnarray}

In contrast, the tree-level calculation of fragmentation  of the heavy 
quark into  hadron $h$, defined as~\cite{Collins:1981uw}:
\begin{eqnarray}
D_{h/Q}(z) = z \int \frac{dy^-}{2\pi} e^{i\frac{p^+}{z}y^-}\frac{1}{3}
{\rm Tr}_{color}\frac{1}{2}{\rm Tr}_{Dirac} \frac{\gamma^+}{2}
\nonumber \\
\times \langle 0 |\psi(y^-,{\bf 0})
a_h^{\dagger}(P^+)a_h(P^+)\bar{\psi}(0,{\bf 0}) |0 \rangle \, ,
\label{hqFF}
\end{eqnarray}
can be shown to be proportional to $\langle 0|  b_q^{\dagger }
( \cdots )  b_q ( \cdots ) | 0 \rangle = 0$ and, hence, vanishes.
This result is also reinforced  by the mismatch at tree-level between  
the large light-cone momentum of the parent heavy quark and the  
corresponding momentum of the heavy quark inside the  meson. 
Since both  $ z=p^+_h / p^+_Q \leq 1$ and  $ x_Q \leq 1 $,   
in the region of interest $D_{h/Q}(0<z<1) \equiv 0$.

\begin{figure}[!b]
\vspace*{.2in}
\includegraphics[width=3.0in,height=1.4in,angle=0]{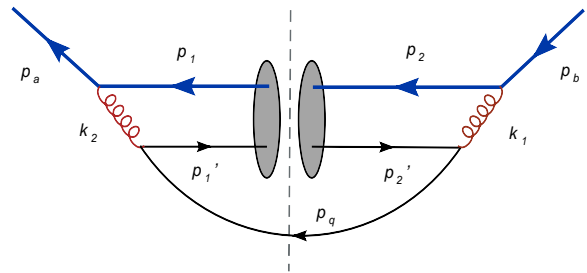} \\
%\vspace*{-1.in}
\caption{ Perturbative QCD calculation of heavy quark fragmentation
matched to the lowest lying Fock state in the heavy meson wavefunction. }
\label{pQCDFF}
\end{figure}

The first non-trivial contribution to $D_{h/Q}(z)$ comes from 
the following
matrix element that replaces the tree-level expression in 
Eq.~(\ref{hqFF}):
\begin{widetext}
\begin{eqnarray}
  && \!\!\!\!\!\! \langle 0 | \psi_Q(y^-,{\bf 0})
\int d^4 y_1 (-ig A_{\mu}^{\alpha}\bar{\psi}_Q\gamma^{\mu}
T^{\alpha} \psi_Q )
\int d^4 y_2 (-ig A_{\nu}^{\alpha}\bar{\psi}_q \gamma^{\nu}
T^{\alpha} \psi_q )  \,  \times \,  a_Q^{\dagger}(p_1) b_{q}^{\dagger}(p_1^{\prime})
b_{q} (p_2^{\prime}) a_Q(p_2) \nonumber \\
  && \!\!\!\!\!\!   \times \,
\int d^4 y_3 (-ig A_{\sigma}^{\beta}\bar{\psi}_q\gamma^{\sigma}
T^{\beta}\psi_q )  \int d^4 y_4 (-ig A_{\lambda}^{\beta}\bar{\psi}_Q
\gamma^{\lambda} T^{\beta}\psi_Q )
\bar{\psi}(0,{\bf 0}) |0 \rangle  \;,  \nonumber \\
\label{contraction}
\end{eqnarray}
\end{widetext}
and corresponds to the process with  Feynman
diagram shown in  Fig.~\ref{pQCDFF}. $T^\alpha$ are the standard Gell-Mann matrices.
Our approach 
is similar to the one outlined in~\cite{Ma:1994zt,Chang:1991bp,Braaten:1994bz} 
but in the calculation we retain 
$\vec{ {q}}$, the  $ Q \bar{q} $ relative momentum distribution, in the 
matrix element. This will allow us to calculate the lowest order correction 
in ${\vec{q}}$ to $D_{h/Q}(z)$.

After converting the integral over the transverse momentum of the outgoing 
free light parton ${ p}_q $ into an integral over the virtuality of the 
heavy quark we obtain the heavy quark fragmentation functions:
\begin{eqnarray}
& & \!\!\!D_{h/Q}(z) =\int \frac{dx_1 d^2{\bf k}_1 \psi(x_1, {\bf k}_1) }
{(2\pi)^32\sqrt{x_1(1-x_1)} }
\frac{dx_2 d^2{\bf k}_2 \psi^*(x_2, {\bf k}_2) }
{(2\pi)^32\sqrt{x_2(1-x_2)} } 
\nonumber \\
& & \!\!\!  \frac{M(j)_{s_1 s_1^{\prime}} }{\sqrt{2}}
\frac{M(j)_{s_2 s_2^{\prime}} }{\sqrt{2}}
\int ds \; \theta \left(s-\frac{m_h^2}{z}-\frac{m_q^2}{1-z} \right)
\frac{\alpha_s^2 C_F^2}{3}   \nonumber \\
& & \!\!\! {\rm Tr} \Big[ \gamma^+ \frac{i}{\gamma \cdot p_a   - m_Q}
\gamma^{\mu} u_{s_1}(p_1) \bar{v}_{s_1^{\prime}} (p_1^{\prime})
\gamma^{\nu} ( \gamma \cdot p_q + m_q)
\gamma^{\sigma} \nonumber \\
& & \!\!\! v_{s_2^{\prime}}(p_2^{\prime}) \bar{u}_{s_2}(p_2) \gamma^{\lambda}
\frac{i}{\gamma \cdot p_b - m_q} \Pi_{\mu\nu}(p_a-p_q)\Pi_{\sigma \lambda}(p_b-p_q)
\Big] \nonumber \\
& &\frac{1}{{\rm Tr}[\gamma^+ (\gamma \cdot p) ] } \, . 
\label{eq:hqFF-2}
\end{eqnarray}
Here, $\Pi_{\mu\nu}(p_a-p_q)=i(-g_{\mu\nu}+(n_\mu k_\nu+n_\nu k_\mu)/(n\cdot k))$ is the gluon propagator,
with $n^\mu=[0,1,{\bf{0}}]$ in light cone coordinates, and $s= p_a^2 = p_b^2$, with $p_a = p_b = p_Q$.
If we define: 
\begin{eqnarray}
& &T(z) =  M(j)_{s_1 s_1^{\prime}} M(j)_{s_2 s_2^{\prime}} \int ds
\; \theta \left(s-\frac{m_h^2}{z}-\frac{m_q^2}{1-z} \right)
 \nonumber \\
& & \!\!\! {\rm Tr} \Big[ \gamma^+ \frac{i}{\gamma \cdot p_a   -
m_Q} \gamma^{\mu} u_{s_1}(p1) \bar{v}_{s_1^{\prime}} (p_1^{\prime})
\gamma^{\nu} ( \gamma \cdot p_q + m_q)
\gamma^{\sigma} \nonumber \\
& & \!\!\! v_{s_2^{\prime}}(p_2^{\prime}) \bar{u}_{s_2}(p_2)
\gamma^{\lambda} \frac{i}{\gamma \cdot p_b - m_q}
\Pi_{\mu\nu}\Pi_{\sigma \lambda}
\Big] \frac{1}{{\rm Tr}[\gamma^+ (\gamma \cdot p) ] } \, , \nonumber \\
\label{eq:hqFF-pQCD}
\end{eqnarray}
the heavy quark fragmentation functions into heavy mesons are
then given by:
\begin{eqnarray}
D_{h/Q}(z) = \int  \frac{d^3 q_1 \;  f({\vec q}_1) }
{(2\pi)^3 }
\frac{d^3 q_2  \; f^*( {\vec q}_2) }
{(2\pi)^3 } \; T(z) \, .
\label{eq:hqFF-3}
\end{eqnarray}
Note that the connection between the instant-form wavefunction  $f({\vec q}_i)$
and the light-cone wavefunction $\psi(x_1, {\bf k}_1)$ is discussed in detail
in Appendix~\ref{Instant}.  Thus, Eq.~(\ref{eq:hqFF-3}) is equivalent to 
Eq.~(\ref{eq:hqFF-2}).

%%%%%%%%%%%%%%%%%%%%%%%%%%%%%%%%%%%%%%%%%%%%%%%
The spin sums for the $ Q \bar{q} $ $^1S_0$ (P) and $^3S_1$ (V)  states can be derived, yielding:
\begin{eqnarray}
& &\sum_{s_1,s_1^{\prime} }  u_{s_1}(p_1) \bar{v}_{s_1^{\prime}}(p_1^{\prime})
M(P)_{s_1 s_1^{\prime}} \nonumber \\
&=& -\frac{ ( {\not\!p_1} + m_Q)(1+\gamma^0)
( {\not\! p_1^{\prime}} + m_q ) \gamma^5 }
{ \sqrt{2(p_1^0 + m_Q)} \sqrt{2(p_1^{\prime 0} + m_q) } } \, ,
\label{eq:spin-sum-1} \\
& &\sum_{s_1,s_1^{\prime} }  u_{s_1}(p_1) \bar{v}_{s_1^{\prime}}(p_1^{\prime})
M(V)_{s_1 s_1^{\prime}} \nonumber \\
&=& \frac{ ( {\not\!p_1} + m_Q)(1+\gamma^0)
( {\not\! p_1^{\prime}} + m_q ) {\not\!\epsilon_p(\lambda)}}
{ \sqrt{2(p_1^0 + m_Q)} \sqrt{2(p_1^{\prime 0} + m_q) } } \, ,
\label{eq:spin-sum-2}
\end{eqnarray}
where $\epsilon_p(\lambda)$ is the 4-component  polarization vector for 
a specific helicity projection $\lambda$ of the $^3S_1$ state.
In deducing the above expressions,  we can first compute the spin sums 
in the rest frame of the heavy meson, then generalize them to 
an arbitrary frame.

Let us first neglect the effects of quark motion on the meson mass 
and assume in the rest frame that $ M = m_Q +
m_q$, $r= m_q/M$, $ 1-r = m_Q/M$. In a Lorentz boosted frame the 
light and heavy quark momenta can be written as:
\begin{eqnarray}
&&  p_1 = (1-r) p + q_{1P} \,, \,\, p_1^{\prime}=r p 
- q_{1P} \, . 
\label{boostap}
\end{eqnarray}
In Eq.~(\ref{boostap}) $q_{iP}$ is the boosted  relative momentum between the quarks 
in the lowest Fock state (${\vec {q}}$ at rest). A classic 
approximation~\cite{Ma:1994zt,Chang:1991bp,Braaten:1994bz}   is to assume that the 
wavefunction $f(\vec q)$ of the heavy meson as a bound state of two quarks 
goes rapidly to zero when $\vec q \neq 0$. The dominant
contribution is, hence,  given by the region $\vec q\sim 0$ and
in this limit the spin sums in Eq.~(\ref{eq:spin-sum-1}) and 
Eq.~(\ref{eq:spin-sum-2}) can be reduced to: 
\begin{eqnarray}
&& \!\!\!\!\!  \sum_{s_1,s_1^{\prime} }  u_{s_1}(p_1)
\bar{v}_{s_1^{\prime}}(p_1^{\prime}) M(P)_{s_1 s_1^{\prime}} =
\frac{\sqrt{m_Q m_q} }{M} \gamma^5 ({\not\!p} -M) \, , 
\nonumber \\
&&\!\!\!\!\!  \sum_{s_1,s_1^{\prime} }  u_{s_1}(p_1)
\bar{v}_{s_1^{\prime}}(p_1^{\prime}) M(V)_{s_1 s_1^{\prime}} =-
\frac{\sqrt{m_Q m_q} }{M} ({\not\!p} +M)
{\not\!\epsilon_p(\lambda)} \, ,
\nonumber \\
\label{eq:spin-sum-reduced}
\end{eqnarray}
which are the same as the ones used in Refs.~\cite{Ma:1994zt,Chang:1991bp,Braaten:1994bz}.

In this paper we employ the complete expressions for the spin sums, 
Eqs.~(\ref{eq:spin-sum-1}) and~(\ref{eq:spin-sum-2}), and calculate $T(z)$ 
defined in Eq.~(\ref{eq:hqFF-pQCD}). This evaluation is technically very complex,
with ${\cal O}(10^4)$ terms,
and requires the use of packages for Dirac algebra manipulation~\cite{Mertig:1990an}.
Having obtained the full result, we can take the ${q} \rightarrow 0$ limit 
to obtain the lowest ${\cal O}(q^0)$ contribution to heavy quark fragmentation
into the $^1S_0$ state:
\begin{eqnarray}
&&\!\!\!\!\! T_P^{(0)}(z) =\nonumber \\
&& \!\!\!\!\! \frac{(z-1)^2 z}{ 3 M^3 r^2 ((r-1)
z+1)^6} \bigg\{ 3 (r-1)^2
(2 (r-1) r+1) z^4  \nonumber \\
&&\!\!\!\!\! + 2 (r-1) (r (18 r-19)+6) z^3+\left(68 r^2-74
r+21\right) z^2 \nonumber \\
&&\!\!\!\!\! + (36 r-18) z+6 \bigg\} \, ,  \label{eq:hqFF-LO}
\end{eqnarray}
and the one into the $^3S_1$ state:
\begin{eqnarray}
&& \!\!\!\!\! T_V^{(0)}(z) =\frac{(z-1)^2 z}{M^3 r^2 ((r-1) z+1)^6} \nonumber \\
&&\!\!\!\!\! \bigg\{z \left((r-1)^2 (2 (r-1) r+3) z^3+2 (r-1)
  (r (2 r-1)+4) z^2 \right.  \nonumber \\ 
&&\!\!\!\!\! \left.  +6 r (2 r-1) z+9 z+4
  r-6\right)+2 \bigg\} \,  ,
\label{eq:triplet-LO}
\end{eqnarray}
respectively, which recover the results given by~\cite{Ma:1994zt,Braaten:1994bz}.

Next, we proceed to evaluate the  ${\cal O}(q^1)$ correction to the fragmentation
function. To do so, we not only select the terms with $1$-st power of $q_{1P}$
and $q_{2P}$, but also take the large $p_Q^+ $ limit (keeping the leading  power of $p_Q^+$). 
We are able for the first time to obtain a meson structure-dependent correction 
to the heavy quark fragmentation functions.  For the $^1S_0$ state or result reads:
\begin{eqnarray}
&&\!\!\!\!\! T_P^{(1)}(z)= \nonumber \\
&&\!\!\!\!\! \frac{(z-1)^2 z^2}{6 M^3 (1-r) r^3 ((r-1)
z+1)^8} \left[\frac{p_Q^+}{q_{1P}^+}\frac{{\bf q}_{1T}^2}{M^2} 
+\frac{p_Q^+}{q_{2P}^+}\frac{{\bf q}_{2T}^2}{M^2}\right] \nonumber \\
&&\!\!\!\!\! \bigg\{ 3 (r-1)^4 r z^6+(r-1)^3 (2 r+5) z^5 + 3 (r-1)^3 (4 r-7)z^4 \nonumber \\
&&\!\!\!\!\! -(r-1) (r (r (8 r-63)+88)-36) z^3 +[r ((81-28 r) r-90)  \nonumber \\
&&\!\!\!\!\!  +32 ] z^2 +((25-8 r) r-15) z+3 \bigg\} \,  ,
\label{eq:hqFF-NLO} \, 
\end{eqnarray}
while for the $^3S_1$ state we get:
\begin{eqnarray}
&&\!\!\!\!\!  T_V^{(1)}(z) = \frac{(z-1)^2 z^2}{6 M^3 (1-r) r^3 ((r-1)
z+1)^8} \nonumber \\
&&\!\!\!\!\!  \bigg\{
\Big(3 z^4 \left(z^2+z+2\right) r^5-z^3 \left(12
  z^3+19 z^2+6 z-16\right) r^4   \nonumber \\
&&\!\!\!\!\! \left. +z^2 \left(18 z^4+50 z^3-68
  z^2+51 z-36\right) r^3  -z \left(12 z^5+66 z^4 \right. \right.  \nonumber \\[1ex]
&&\!\!\!\!\!  \left. \left.  -176 z^3+203
  z^2-121 z+16\right) r^2   +(z-1)^2 \left(3 z^4+49 z^3  \right.      \right.  \nonumber \\
&& \!\!\!\!\! \left. \left. -59 z^2+50
  z-6\right) r \right.  -(z-1)^3 \left(11 z^2-13 z+9\right)\Big) 
\frac{p_Q^+}{q_{1P}^+}\frac{{\bf q}_{1T}^2}{M^2}  \nonumber \\
%**********************
&&\!\!\!\!\! + \Big(3 r^6 z^6+r^5 \left(-6 z^2+13 z+14\right)
  z^4  -r^4 \left(12 z^3+27 z^2   \right.  \nonumber \\
&&\!\!\!\!\! \left. \left. +29 z-32\right) z^3 \right. 
%\nonumber \\ &&\!\!\!\!\! 
\left.+r^3 \left(48
  z^4-16 z^3+20 z^2-9 z-28\right) z^2 \right. \nonumber \\[1ex]
&&\!\!\!\!\! \left. -r^2 \left(57 z^5-80
  z^4+26 z^3+57 z^2-76 z+16\right) z \right. \nonumber \\[1ex]
&&\!\!\!\!\! \left.-(z-1)^3 \left(6 z^3-z^2-4
  z+6\right) \right. +r (z-1)^2 \left(30 z^4-9 z^3 \right. \nonumber \\
&&\!\!\!\!\!  \left. -10 z^2+32
  z-6\right)\Big) 
\frac{p_Q^+}{q_{2P}^+}\frac{{\bf q}_{2T}^2}{M^2} \Bigg\}  \; .
\label{eq:triplet-NLO} 
\end{eqnarray}
In Eq.~(\ref{eq:triplet-NLO}) we observe that for the triplet state 
the coefficients of $q_{1P}$
and $q_{2P}$ are different, which results from the different
combinations of $\gamma$ matrices. For example, a typical term 
in the intermediate results for the $^3S_1$ state may look like
$${\rm Tr}\big[A {\not\!q_{1P}}(1+\gamma^0)({\not\!p}+M)B 
({\not\!p}+M)\big] \, ,$$  where A and B are  polynomials of Dirac matrices.  
This is not always equal to:
$${\rm Tr}\big[A ({\not\!p}+M)B ({\not\!p}+M)(1+\gamma^0)
{\not\!q_{2P}} \big] \; ,$$
even after exchanging the momenta $(1\leftrightarrow 2)$. 
On the other hand, $ q_{i\, T}, \, q^+_{i\, P}$ (or $\vec{q}$ in the rest frame) 
are themselves internal integration variables with equal weights
$\psi(x_i,{\bf k_i})$  (or $f(\vec{q}_i)$ in instant form), see 
Eqs.~(\ref{eq:hqFF-2}) and (\ref{eq:hqFF-3}). Hence, 
Eq.~(\ref{eq:triplet-NLO}) can be written in a symmetric form if
desired. 
%This symmetrization is also facilitated by the real wavefunctions.

%even if we only keep the leading term in $p_Q^+$. Here
%both $A$ and $B$ are combinations of five Dirac $\gamma$ matrices
%with some prefactors.
%However, the above expression can always be  
%written in a symmetrical form taking into account that
%in the last step to derive
%heavy meson fragmentation functions we should integrate out $q_{1}$ and 
%$q_{2}$, and utilizing the relation
%\begin{eqnarray}
%&&\int  \frac{d^3 q_1 \;  f({\vec q}_1) }{(2\pi)^3 }
%\frac{d^3 q_2  \; f^*( \vec{q}_2) }{(2\pi)^3 } g( q_1)
%\nonumber \\
%&=&\int  \frac{d^3 q_1 \;  f({\vec q}_1) }{(2\pi)^3 }
%\frac{d^3 q_2  \; f^*( {\vec q}_2) }{(2\pi)^3 } g(q_2) \, ,
%\end{eqnarray}
%where $g(q_1)$(or $g(q_2)$) is any function of $q_1$(or $q_2$)
%and the heavy meson wavefunction $f^*({\vec q}) =  f({\vec q})$.

Finally, we note that if the relative momentum of the light and heavy quarks
is not neglected then in a boosted frame the large momentum of the meson in
Eq.~(\ref{boostap}) will be partitioned proportional to the parton transverse mass, 
$ m_{T\, Q} = \sqrt{m_Q^2+\langle {\bf k}^2 \rangle}, \; 
  m_{T\, q} = \sqrt{m_q^2+\langle {\bf k}^2 \rangle} .$ In this more general case
$M = m_{T\, Q} + m_{T\, q}$ and $r =  m_{T\, q} / M$, $1-r =  m_{T\, Q} / M$. 
In this work  we calculate  $\langle {\bf k}^2 \rangle = 
\langle {\bf q}^2 \rangle$ from the light-cone wavefunctions of heavy mesons
and use the corresponding values of $r$ both in the vacuum and at $T\neq 0$. 
This allows for a more consistent description of heavy quark fragmentation 
without treating  $r$ as a phenomenological  parameter.

\begin{figure}[!t]
\vspace*{.2in}
\includegraphics[width=3.2in,height=2.6in,angle=0]{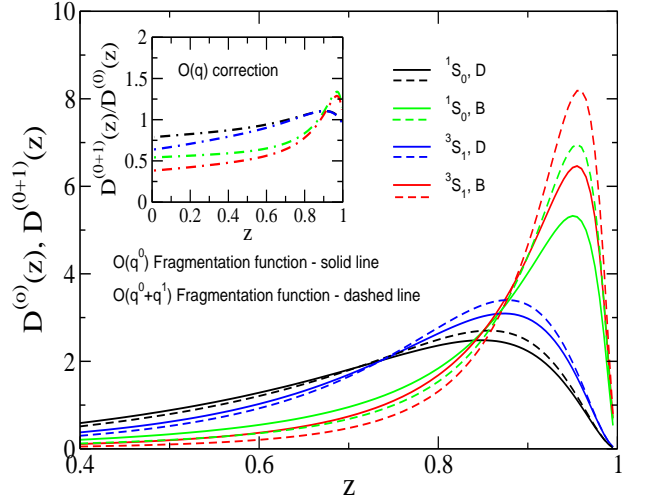} 
%\vspace*{-1.in}
\caption{ Fragmentation functions of charm and beauty quarks into $^1S_0$ and  $^3S_1$  
 heavy meson states. Solid lines are for negligible momentum spread  between the 
parton constituents of  $D$ and $B$. Dashed lines represent the first calculation 
for $\vec{q} \neq 0$. Insert shows the fractional correction that arises from the 
structure of the final-state hadrons for both the pseudoscalar and vector states. } 
 \label{pQCDFF-fig}
\end{figure}

With expressions for $T(z)$ at hand, after integrating over 
the relative momenta of $Q{\bar q}$ system in the rest frame of  the
heavy meson, we obtain the $z$-dependence of heavy quark fragmentation
functions for the $^1S_0$ and $^3S_1$ states. Keeping in mind that 
the relative momentum is rather small and in the calculation $|q|\ll rM$ and
$|q|\ll (1-r)M$ was implicit, we choose
the  upper bound of this integration to be $rM/2$ 
due to the fact that $r<(1-r)$. The absolute normalization of 
$D_{h/Q}(z)$ has been fixed by the known branching ratios of the charm
and beauty quarks into heavy mesons and baryons, such that 
$\sum _i\int_0^1 \, D_{h_i /Q}(z) dz = 1.$

%%%%%%%%%%%%%%%%%%%%%%%%%%%%%%%%%%%%%%%%%%%%%%%%%%%%%%%%%

Numerical results for $c$ and $b$ quark fragmentation into  the $^1S_0$ and the $^3S_1$  
$D$ and $B$ mesons states, respectively, are shown in  Fig.~\ref{pQCDFF-fig}. 
The solid lines in the figure denote  the standard ${\cal O}(q^0)$ calculation of $D_{h/Q}(z)$, 
in accord with~\cite{Ma:1994zt,Braaten:1994bz}. The dashed lines signify the addition of the  
${\cal O}(q^1)$ correction, the first correction that arises from the internal
structure of the final-state meson. The magnitude of the  ${\cal O}(q^1)$ 
term is best seen in the insert of Fig.~\ref{pQCDFF-fig} in the ratio
$D^{(0+1)}(z) / D^0(z)$. While on an absolute scale the correction to the 
shape of heavy quark fragmentation can be $ \pm 50\% $, it doesn't change appreciably
the position of the large-$z$ peak and, hence, will not alter the cross section
for the observed $D$ and $B$ meson production. To facilitate the  comparison between 
the FFs in the vacuum and at $T \neq 0$ we will only use the leading result for structureless
hadrons, Eqs.~(\ref{eq:hqFF-LO}) and (\ref{eq:triplet-LO}).

%%%%%%%%%%%%%%%%%%%%%%%%%%%%%%%%%%%%%%%%%%%%%%%%%%%%%%%%%%%%%%%%%%%%%%%%%%%%

\section{The fate of  heavy mesons below and above $T_c$}
\label{potmod}
In this section we move beyond the vacuum $T=0$ case.
Potential models have been quite useful in studying the non-perturbative aspects of
QCD. In particular, for quarkonia one can put the potential model on firm
grounds by using an effective field theory (NRQCD)~\cite{Brambilla:2004jw}. More
generally, heavy-light mesons like the $D$ and the $B$ and their excited states at
zero temperature have been studied in detail~\cite{Avila:1994vi,Avila:1999aj} by
treating the light  $\bar{q}$ as a Dirac particle moving in a confining
potential set up by the heavy quark $Q$. In this section we investigate the
bound-state wavefunctions for $D$- and $B$-mesons at
rest and in equilibrium with a thermal medium or in a co-moving plasma.  
%If we assume that a meson moving with
%respect to the medium is in thermal equilibrium with the medium, and hence its
%wavefunction is simply related to the wavefunction at rest by a boost,
These can be employed to calculate the PDFs and FFs for quarks/mesons, as discussed in
Appendix~\ref{Instant}. 

Lattice QCD results for the free energy of a system at temperature $T$ containing
two infinitely heavy quarks separated by a distance $r$~\cite{Kaczmarek:2005ui} have
been used to extract the static potential between these quarks: 
\begin{equation}
V(r)=\Biggl\{
\begin{array}{cc}
  -\frac{\alpha}{r}+\sigma r \; ,  & r<r_{med}(T)\\[2ex]
  -\frac{\alpha_1(T)\exp(-\mu(T) r)}{r}+\sigma r_{med}(T) \; , &
  r>r_{med}(T)
\end{array} \label{pf}\;  . 
\end{equation}
Here, $r_{med}(T)$ signifies the temperature-dependent distance
scale at which the medium breaks off the linear confining potential, $\alpha_1(T)$
represents the effective coupling at a given temperature, 
and $\mu(T)$ is the Debye screening mass. Interpolation is
typically used to smooth out the sharpness in the potential at $r=r_{med}$ and
we describe the interpolation below.
This potential has been employed previously to study the fate of heavy-heavy
bound states in the QGP formed at temperatures above
$T_c$~\cite{Mocsy:2007yj,Mocsy:2007jz}.

For future convenience, it is useful to separate the ``vector'' ($V_v(r)$) and
``scalar'' ($V_s(r)$) parts of the potential in Eq.~(\ref{pf}). This is done as
described for $T=0$ in~\cite{Avila:1994vi,Avila:1999aj}.  In Eq.~(\ref{pf}) the
term  proportional to $\sigma$, which in the absence of a medium would be a
confining linear potential, is taken to be scalar. The Coulomb part is naturally
interpreted as the zeroth component of a vector potential. Thus we consider:
\begin{equation}
\begin{split}
V_v(r)&=\bigl\{
\begin{array}{cc}
  -\frac{\alpha}{r} & r<r_{med}\\
  -\frac{\alpha_1\exp(-\mu r)}{r} & r>r_{med}
\end{array} \; ,  \\ 
V_s(r)&=\bigl\{
\begin{array}{cc}
  \sigma r & r<r_{med}\\
  \sigma r_{med} & r>r_{med}
\end{array}  \; . \label{separated potential form}\;  
\end{split} 
\end{equation}

The typical parameters used in Eq.~(\ref{pf}) are {$\alpha=0.445$},
$\alpha_1(T)=\frac{1}{3\pi} g^2(2\pi T)$, and $\mu(T)=1.417
\sqrt{1+\frac{N_f}{6}}g(2\pi T) T$.  For $T>1.1T_c$, we use the perturbative two
loop expression for $g(2\pi T)$. For $T$ below $0.9T_c$, where the perturbative
expression for $g(2\pi T)$ does not work, we take $g = 2.05$ which corresponds
to $\alpha=0.445$, and near $T_c$ we interpolate between the two. The strength
of the confining potential is taken to be $\sigma=0.224$~GeV$^2$ and $T_c =
192$~MeV.  The value of  $\sigma$ is $\approx 5$\% larger than
in~\cite{Mocsy:2007jz} and we take $M_c=1.34$~GeV and $M_b=4.79$~GeV.
$r_{med}=\min (0.4\frac{T_c}{T},1.1)$~fm, where $1.1$~fm is the typical string
breaking scale~\cite{Mocsy:2007jz}. $V_s(r)$ is smoothed out on a scale much
smaller than $r_{med}(T)$. $V_v(r)$, following~\cite{Mocsy:2007jz}, is smoothed
by interpolating between $\alpha/r$ at $r=r_{med}(T)$ and $\alpha_1\exp(-\mu
r)/r$ at $r=r_1(T)$, where $r_1(T)$ is larger than $r_{med}(T)$. For our
calculations, we take $r_1(T)=2r_{med}(T)$ for $T<1.1T_c$ and $r_1(T)=1.25$~fm for $T>1.1T_c$.
This gives a slightly stronger potential than that used by~\cite{Mocsy:2007jz}.   

The presence of bound states of heavy-heavy systems, like charmonia and
bottomonia, can be explored by solving the non-relativistic
Schr$\ddot{{\rm{o}}}$dinger equation for the heavy quarks in the presence of the
potential Eq.~(\ref{pf}). Here, we only look at the $l=0$ wavefunction, for which the
equation has a form,
\begin{equation}
\frac{1}{2(m_Q/2)}\frac{1}{r}\frac{\partial^2 r\psi}{\partial r^2} + V(r)\psi =
E\psi(r)\label{schroedinger}\;,
\end{equation}
where $m_Q$ is the mass of the heavy quark. We find that $J/\psi$ bound state solutions exist up
to about $2 T_c$ though they have a binding energy, quite sensitive to the
precise value of $\sigma$, much less than the temperature. The $\Upsilon$
disappears at much higher temperatures but its binding energy  becomes smaller
than $T$  near $1.5 T_c$. These values are larger than the values reported
in~\cite{Mocsy:2007jz} because of two reasons. First, the value of $\sigma$ is a
little larger. Second, we use a slightly different interpolation to smooth the
potential between $r<r_{med}$ and $r>r_{med}$. Table~\ref{QQ} shows the binding
energy and sizes for the $J/\psi$ and the $\Upsilon$ at various temperatures
above $T_c$.
\begin{table}[!b]
\begin{tabular}{c|c|c|c}\\
\ \ $T/T_c$\ \  & \ \ $T$ (GeV) \ \ & \ \ $E_b$ (GeV) \ \ & \ \ $\langle r\rangle$ (fm) \ \ \\
\hline
$1.2$ & $0.230$ & $0.042$ & $0.722$\\
$1.4$ & $0.269$ & $0.030$ & $0.804$\\
$1.6$ & $0.307$ & $0.024$ & $0.869$\\
$1.8$ & $0.346$ & $0.020$ & $0.923$\\
$2.0$ & $0.384$ & $0.017$ & $0.968$\\
\hline\hline
$1.2$ & $0.230$ & $0.344$ & $0.222$\\
$1.4$ & $0.269$ & $0.301$ & $0.232$\\
$1.6$ & $0.307$ & $0.273$ & $0.241$\\
$1.8$ & $0.346$ & $0.254$ & $0.248$\\
$2.0$ & $0.384$ & $0.241$ & $0.254$\\
\hline
\end{tabular}
\caption{The binding energy and mean radius for  $J/\psi$ (upper
table) and $\Upsilon$ (lower table) as a function of the temperature.}
\label{QQ}
\end{table}

To investigate the existence of heavy-light mesons in the QGP we need to solve
the Dirac equation for a light particle moving in the potential created by a
heavy quark:
\begin{equation}
H\psi=E\psi\;,
\end{equation}
where $\psi$ is the four-component Dirac spinor and $H$ is the Dirac
Hamiltonian. $H$ has the following form:
\begin{equation}
H=-i\gamma^0\gamma^i\partial_i+V_v(r)+\gamma^0(m+V_s(r))\;.
\end{equation}
The eigenfunctions of the spherically symmetric Dirac Hamiltonian can be 
written as~\cite{Sakurai,Greiner,Avila:1994vi,Avila:1999aj}:
\begin{equation}
\psi({\bf{r}})= \frac{1}{r}\left(\begin{array}{cc}
  G(r)\\
  i{\bf{\sigma\cdot\hat{r}}}F(r)
\end{array}\right){\cal{Y}}^{j_3}_{jls}\;,
\end{equation}
where ${\cal{Y}}^{j_3}_{jls}$ are angular functions that are obtained by the angular
momentum addition of $s=\frac{1}{2}$ to $l$ giving $j$. 
The energy eigenequations for the radial wavefunctions $F(r)$ and
$G(r)$ that we need to solve are as follows:
\begin{equation}
\begin{split} \left| \begin{array}{ll}
F'(r) - \frac{\kappa}{r}F(r) &= (-E+V_v(r)+m+V_s(r)) G(r)\\[2ex]
G'(r) + \frac{\kappa}{r}G(r) &= (E-V_v(r)+m+V_s(r)) F(r)\label{rde}\;,
\end{array} \right .
\end{split}
\end{equation}
where $\kappa=\pm(j+1/2)$. For the lowest energy states that we are interested
in, $j=1/2$ and $\kappa=-1$. We solve Eq.~(\ref{rde})  numerically
using the ``shooting method''~\cite{press}. Since the potential tends to a
constant at large distances, we begin from an exponentially decaying solution at
a very large distance. We solve the differential equation implied by the Dirac
equation for this boundary condition and choose the value of the energy $E$ such
that the value of the wavefunctions $F$ and $G$ at $r=0$ is $0$.

\begin{table}[!b]
\begin{tabular}{c|c|c|c}\\
\ \ $T$ ($T_c$) \ \ & \ \ $T$ (GeV) \ \ & \ \ $E_b$ (GeV) \ \ & $ \ \ 
\sqrt{\langle r^2\rangle}$ (fm) \ \ \\
\hline
$0$      & $0$     & $0.730$ & $0.468$\\
$0.2T_c$ & $0.038$ & $0.733$ & $0.466$\\
$0.4T_c$ & $0.077$ & $0.611$ & $0.464$\\
$0.6T_c$ & $0.115$ & $0.256$ & $0.501$\\
$0.8T_c$ & $0.154$ & $0.098$ & $0.632$\\
$1.0T_c$ & $0.211$ & $0.043$ & $0.785$\\
$1.2T_c$ & $0.230$ & $0.031$ & $0.970$\\
$1.4T_c$ & $0.269$ & $0.017$ & $1.263$\\
$1.6T_c$ & $0.307$ & $0.009$ & $1.636$\\
\hline
\end{tabular}
\caption{ Properties of the  $D$ and $B$ meson bound-state solutions taking the effective light 
quark mass to be $m(T)/\sqrt{2}$. These persist up to temperatures $\approx 1.6 T_c$. }
\label{mqTby2} 
\end{table}

Since the heavy-light mesons are larger in size when compared to quarkonia, it
is natural to expect that they will be affected more severely by color screening.
Indeed, for a light current quark mass of about $0.005$~GeV, we don't find even 
very weakly bound states for $T>T_c$. However, this
conclusion is quite sensitive to the parameters chosen above. For example,
by simply changing $r_{med}$ from $0.4\frac{T}{T_c}$~fm to
$0.45\frac{T}{T_c}$~fm~\cite{Mocsy:2007yj} we find bound
states up to $ \sim 1.2 T_c$, albeit with binding energies $E_b$ about $50$ times 
smaller than $T$. A quark moving in the QGP will have an
effective thermal mass larger than its bare mass $m_q$. For $T>T_c$ perturbative
calculations give $m(T) = \sqrt{\frac{C_F}{4}\bigl[g(2\pi T)T\bigr]^2+m^2_q }$. 
For $T<T_c$ this form is not valid and we simply interpolate the perturbative
value above $T_c$ to the bare mass at $T=0$. 
One would expect the effective mass of a quark moving under the influence 
of a potential in the thermal medium to lie somewhere between $m_q$ and $m(T)$. 
For example,  results obtained by solving the Dirac equation for the light quark 
mass taken to be $m(T)/\sqrt{2}$ are given in Table~\ref{mqTby2}. 
We remark that the very small rise in the binding energy that can be seen in
Table~\ref{mqTby2} as we increase the temperature from  $T=0$ to $0.2T_c$ 
is an artifact of the interpolation we have used for the
thermal mass below $T_c$ and is not numerically significant to affect our results.
The use of  light quark mass equal to
$m(T)$ gives bound-state solutions with weak binding energies up to even above 
$2T_c$, see Table~\ref{mqT}. We also establish the existence of
$D$ and $B$ meson bound-state solutions well above the phase transition temperature for
constituent light quark masses and/or for the ``maximum confining potential''
compatible with lattice data: all the way up to $\sim 2T_c$ for $m_q =
m(T)/\sqrt{2}$ and $\sim 3T_c$ for $m_q = m(T)$.

\begin{table}[!t]
\begin{tabular}{c|c|c|c}\\
\ \ $T \; (T_c)$ \ \ & \ \ $T$ (GeV) \ \ & \ \ $E_b$ (GeV) \ \ & \ \ 
$\sqrt{\langle r^2\rangle}$ (fm) \ \  \\
\hline
$\cdots$ & $\cdots$ & $\cdots$ & $\cdots$ \\
$1.0T_c$ & $0.211$ & $0.060$ & $0.670$\\
$1.2T_c$ & $0.230$ & $0.047$ & $0.797$\\
$1.4T_c$ & $0.269$ & $0.032$ & $0.933$\\
$1.6T_c$ & $0.307$ & $0.024$ & $1.044$\\
$1.8T_c$ & $0.346$ & $0.020$ & $1.119$\\
$2.0T_c$ & $0.384$ & $0.018$ & $1.158$\\
\hline
\end{tabular}
\caption{ Properties of the bound $D$ and $B$ meson states taking the effective light 
quark mass to be $m(T)$. }
\label{mqT}
\end{table}

\begin{figure}[!t]
\vspace*{.0in}
\includegraphics[width=3.0in,height=3.4in,angle=0]{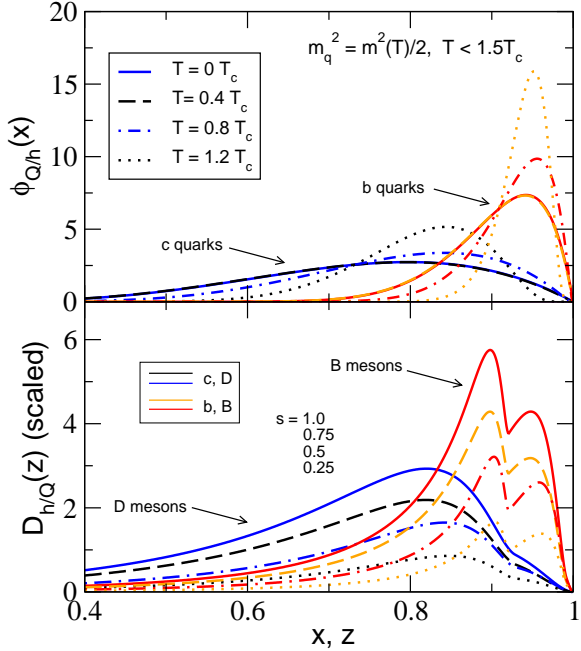} 
\caption{ The evolution of PDFs $\phi_{Q/H}(x)$ (top panel) and the 
FFs $D_{H/Q}(z)$ (bottom panel)  with $T$. Blue/black lines refer to charm 
while  red/orange lines refer to beauty. A scale factor $s$ has been 
used to better separate the decay probabilities. }
\label{pdfff}
\end{figure}

The bound-state wavefunctions, calculated at  temperatures  $T=0T_c, \;  0.4T_c,\;  
0.8T_c,\;  1.2 T_c$ with the light quark mass taken to be $m(T)/\sqrt{2}$ are  employed 
to evaluate the PDFs and FFs of heavy quarks in a 
co-moving  thermal  medium from Eqs.~(\ref{PDFdef2}) and (\ref{eq:hqFF-2}) 
and our results are presented in  Fig.~\ref{pdfff}. The distribution 
functions $\phi_{Q/H}(x)$ become narrower in $x$ as we increase
the temperature, which is intuitively expected  because the wavefunctions become
broader in position space. In contrast the peak position and  width of the decay 
probabilities are determined by the boost parameter $r$. These remain nearly
constant because the decrease in the mean transverse momentum squared is largely 
compensated  by the growth in the thermal mass. Hence, we have used an arbitrary 
scale factor $s$ to better separate the curves in the bottom panel of 
Fig.~\ref{pdfff}. We remark that the double-peak structure  arises from the decay
of the excited $^3S_1$ states into $^1S_0$ states which effectively shifts the
momentum fraction $z$ to lower values.

To summarize, as a rule we find that $D$ and $B$ meson  bound-state
solutions persist above $T_c$. Their small binding energy and large radius, however, will 
greatly facilitate their subsequent dissolution in the presence of interactions.

%%%%%%%%%%%%%%%%%%%%%%%%%%%%%%%%%%%%%%%%%%%%%%%%%%%%%%%%%%%%%%%%%%%%%%%%%%%%
\section{Application to heavy meson production in heavy ion collisions}
\label{application}

Heavy flavor dynamics in dense QCD matter critically depends on the time  scales 
involved in the underlying reaction. Two of these timescales, $\tau_0$ and $L_{QGP}$,  
can be related to the nuclear geometry, the QGP expansion,  and the properties of 
bulk particle production~\cite{Vitev:2008rz}. They signify the onset and 
disappearance of QGP-induced effects and determine, in relation to the formation 
time  Eq.~(\ref{tfrag}), whether the suppression of the observed cross section 
arises predominantly at the hadronic ($\tau_{\rm{form}} \ll L_{QGP}$) or 
partonic ($\tau_{\rm{form}} \geq L_{QGP} $) level.  As discussed 
in the previous section, the dissociation of $D$ and $B$
mesons in the vicinity of $T_c$ can be facilitated by their  small binding 
energy, or, equivalently, their broad wavefunction in coordinate
space. Whether such thermal effects take place in practice, however, depends 
on the time they need to develop. We can roughly estimate this time by 
boosting  the expanded size of the hadron,  $\approx 2 \sqrt{\langle r^2 \rangle}  $ 
from  Table~\ref{mqTby2}, by the $\gamma=1/\sqrt{1-v^2}$  factor. When compared to  
$\tau_{\rm{form}} $, which is determined by the virtuality in the parton
decay process, this time is large and suggests that the fragmentation 
component of the heavy meson dynamics in heavy ion collisions may not be 
affected by the QGP. In what follows we will study this ``instant wavefunction 
limit'' in detail. Before we proceed we remark that in Appendix~\ref{pickup} 
we have also estimated  the correction to the FF $D_{h/Q}(z)$ from the 
pick-up of a light thermal quark strictly in the fragmentation region $0< z < 1$ 
and found this effect to be small,  $\leq 3\% $ and $15\%$ for $D$ and 
$B$ mesons at $p_T \geq 4$~GeV, respectively.

A meson that is formed and propagates inside the medium will undergo collisional 
broadening and dissociation~\cite{Adil:2006ra,Dominguez:2008be}. The dissociation 
rate can be evaluated as follows: 
\begin{eqnarray}
\frac{1}{\langle \tau_{\rm diss}(p_T, t) \rangle} &=& 
\frac{\partial}{\partial t} \ln (1- P_s(p_T,m_Q,t)  ) \; ,  
\label{avtdiss}
\end{eqnarray} 
where the survival probability is given by the overlap of the initial wavefunction $\psi_0$
with the wavefunction at time $t$, $\psi_f$: 
\begin{eqnarray}
&& \hspace*{-0.7cm} P_s \left(\frac {\mu^2 L}{\lambda_q} 
\xi\right) 
= \left| \int d^{2} {\bf k} dx \,
\psi_{f}^* ( {\bf k},x)\psi_{0}( {\bf k}, x) \right|^{2} \;,
\label{eq:survPro}
\end{eqnarray}
Details of the diagrammatic calculation and resummation of the interactions 
that lead to the final-state broadening of the $D$ and $B$ meson wavefunctions, 
$\psi_{0}( {\bf k}, x) \rightarrow \psi_{f} ( {\bf k},x) $,  can be found in~\cite{Adil:2006ra}. 
We here remark that the harmonic oscillator wavefunction form, discussed in 
Appendix~\ref{Instant}, allows for an analytic evaluation of this broadening.
As in Eq.~(\ref{Cron}),  $\frac {\mu^2 L}{\lambda_q}  \xi$ is proportional to the
cumulative transverse momentum transfer from the medium to the propagating system 
and is here calculated for Bjorken-expanding QGP. Formation rates are calculated by
taking into account all final-state heavy meson states   
\begin{eqnarray} 
\frac{1}{\langle \tau_{\rm form}(p_T, t) \rangle } &=& 
\bigg[\sum_i  \int_0^1 dz \, D_{H_i/Q}(z)  \nonumber \\
&& \times \tau^i_{\rm form}(z,p_T,m_Q, t) \, \bigg]^{-1}  \;.
\label{avtform}
\end{eqnarray}

\begin{figure}[!t]
\vspace*{.2in}
\includegraphics[width=3.0in,height=3.4in,angle=0]{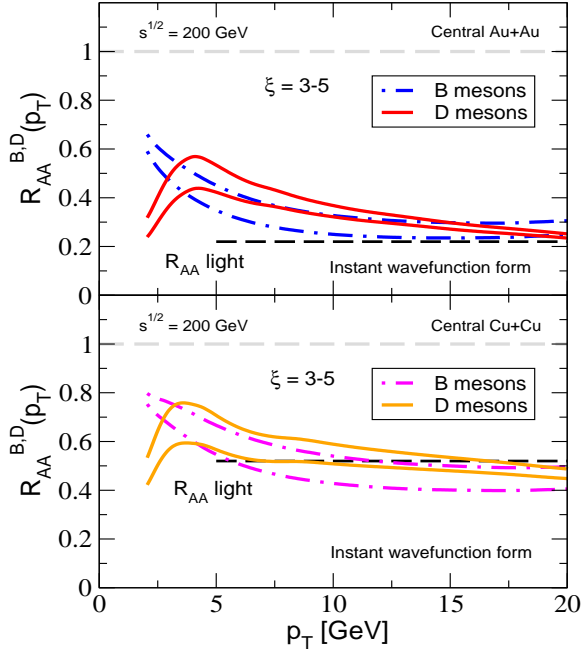}
\caption{ Suppression of $D$ and $B$ hadron production from  meson
dissociation and heavy quark quenching in central Au+Au (top panel)
and Cu+Cu collisions (bottom panel) at $ \sqrt{s_{NN}} =200$~GeV at
RHIC. } 
\label{fig:BDSupp-RHIC}
\end{figure}

%%%%%%%%%%%
Employing the above calculated rates, Eqs.~(\ref{avtdiss}) and (\ref{avtform}),
the concurrent processes of $c$ and $b$ quark fragmentation and $D$ 
and $B$ meson dissociation are described by the following set of 
rate equations~\cite{Adil:2006ra}:  
\begin{eqnarray}
\label{rateq1}
\partial_t f^{Q}({p}_{T},t) &=&
-  \frac{1}{\langle \tau_{\rm form}(p_T, t) \rangle} f^{Q}({p}_{T},t)
\nonumber \\
&& \hspace*{-2cm}  + \, \frac{1}{\langle
\tau_{\rm diss}(p_T/\bar{x}, t) \rangle}
\int_0^1 dx \,  \frac{1}{x^2} \phi_{Q/H}(x)
f^{H}({p}_{T}/x,t) \;, \qquad \\[1ex]
\partial_t f^{H}({p}_{T},t) &=&
-  \frac{1}{\langle \tau_{\rm diss}(p_T, t) \rangle} f^{H}({p}_{T},t)
\nonumber \\  
&& \hspace*{-2cm} +\, \frac{1}{\langle
\tau_{\rm form}(p_T/\bar{z}, t) \rangle}
\int_0^1 dz \,  \frac{1}{z^2} D_{H/Q}(z)
f^{Q}({p}_{T}/z,t) \; . \qquad
\label{rateq2}
\end{eqnarray}
Here, $f^{Q}(p_T,t)$ ($f^{H}(p_T,t)$) is the differential cross section to find a 
quark $Q$ (hadron $H$) with a fixed
rapidity $y$ and transverse momentum $p_T$, at a time $t$. We have suppressed the rapidity
dependence for clarity of notation. Note that the reason for which the
asymptotic $t \rightarrow \infty$ solution will exhibit suppression of the cross 
sections is that
both fragmentation and dissociation processes emulate energy loss by shifting 
the quarks/hadrons to
lower momenta. Under certain simplifying assumptions the system of equations, 
Eqs.~(\ref{rateq1})
and~(\ref{rateq2}), can be solved analytically, as shown in Appendix~\ref{DiffEq}.

At present, there is no reliable way of incorporating the fluctuations
in partonic energy loss in rate or transport equations. Therefore, we include 
the early-time heavy quark inelastic scattering effects approximately as a 
quenched initial condition:
\begin{equation}
\left| \begin{array}{rcl}
f^{Q}({p}_{T},t)&=& \frac{d\sigma^Q(t)}{dy d^2{\bf{p}}} \;,
\;   f^{Q}({p}_{T},0) =
\frac{d \sigma^{Q,{\rm Quench}}}{dy d^2{\bf{p}}} \;, \qquad
\\[2ex]
f^{H}({p}_{T},t)&=& \frac{d\sigma^H(t)}{dyd^2{\bf{p}}} \;,
\;   f^{H}({p}_{T},0) = 0 \; .
\end{array}
\right. 
\label{eq:initial-condition}
\end{equation}
Here, the attenuated partonic spectrum 
$\frac{d \sigma^{Q,{\rm Quench}}}{dy d^2{\bf{p}}}$  
is calculated differentially versus $p_T$ using Eq.~(\ref{Quench}). The relevant 
mean quenching time - the time that the  physical system of interest spends in a quark state - can be 
estimated from the analytic solution for $f^{Q}({p}_{T},t), \, f^{H}({p}_{T},t)$  given
in Appendix~\ref{DiffEq}. On the other hand, $\tau_{\rm form}$  represents   the important
stage until the first fragmentation of the heavy quark into heavy meson. We use the 
average of these two results to calculate the initial condition, Eq.~(\ref{eq:initial-condition}).

\begin{figure}[!t]
\vspace*{.2in}
\includegraphics[width=3.0in,height=2.5in,angle=0]{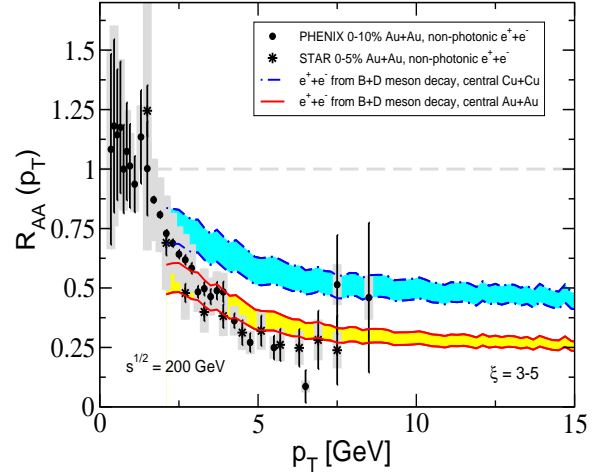}
\caption{Nuclear modification for the single non-photonic electrons
in central Au+Au and Cu+Cu collisions at RHIC.  Data is from
PHENIX~\cite{Adare:2006nq} and STAR~\cite{Abelev:2006db}
collaborations.} \label{fig:eSupp-RHIC}
\end{figure}

%%%%%%%%%%%%%%%%%%%%%%%%%%%%%%%%%%%%%%%%%%%%%%%%%%%%%%%%%%%%%

We integrate numerically the above set of coupled ordinary differential equations 
and  use the same initial soft gluon rapidity  density $dN^g/dy$ as in the 
simulations of $\pi^0$ quenching. The corresponding  suppression of open
heavy flavor in $\sqrt{s_{NN}} =200$~GeV central Au+Au and Cu+Cu collisions 
at RHIC is shown in the top and bottom panels of Fig.~\ref{fig:BDSupp-RHIC}, 
respectively. For $D$ mesons the Cronin effect is clearly visible around 
$p_T \sim 4$~GeV and this is a notable difference from our previous 
study~\cite{Adil:2006ra} where initial-state $k_T$ diffusion was not included. 
For $B$ mesons the change in heavy quark velocity $\beta_Q$, which controls 
the $Q\bar{q}$ ($\bar{Q} q $) broadening  $\propto \beta_Q \frac{\mu^2_0}{\lambda_0} 
\xi \ln \frac{\tau}{\tau_0} $ and dissociation, results in the
characteristic rapid decrease in $R_{AA}^{B}(p_T)$ in this part of phase space.   
In both gold and copper reactions at RHIC the suppression  $R_{AA}^B \approx   
R_{AA}^D$  for $p_T > 4$~GeV and these approach the quenching of light hadrons
for $p_T > 10$~GeV. It should be noted that while for the $D$ mesons we observe
a transition from collisional dissociation to partonic energy loss in the studied
kinematic domain, for B meson at RHIC the competing hadronic processes, 
Eqs.~(\ref{rateq1}) and (\ref{rateq2}), still play the dominant role.

Our theoretical results, presented in Fig.~\ref{fig:BDSupp-RHIC},  are 
most relevant to the future vertex detector upgrades at RHIC that will 
ensure direct and separate measurements of the $D$ and $B$ mesons. These 
new experimental data will then allow to  pinpoint the mechanisms of
heavy flavor suppression in the QGP. Indirect studies  of $D$ 
and $B$ meson attenuation are currently carried out through the semi-leptonic 
decays of the charm and beauty hadrons.  We use the PYTHIA event 
generator~\cite{Sjostrand:2006za} to simulate the full kinematics of
these Dalitz decays in p+p and A+A reactions. The nuclear 
modification ratio $R_{AA}^e (p_T)$ of inclusive non-photonic electrons 
is presented in Fig.~\ref{fig:eSupp-RHIC} for $\sqrt{s_{NN}}=200$~GeV
collisions at RHIC. Predictions for central Cu+Cu collisions are also shown 
for  comparison to upcoming STAR data. We remark that the Cronin effect
included in our study changes the absolute scale of the differential
heavy quark cross section at intermediate $p_T$.  Consequently, to  
obtain a good  description of the  non-photonic $e^+ + e^-$ quenching 
results from  PHENIX~\cite{Adare:2006nq} and  STAR~\cite{Abelev:2006db} we 
use $\xi = 3-5$,  approximately 50\%  larger than in our previous 
study~\cite{Adil:2006ra} but still compatible with the enhancement of 
the parton broadening that  comes from the power-law momentum transfer 
tails of the in-medium Moliere scattering.

\begin{figure}[!t]
\vspace*{.2in}
\includegraphics[width=3.0in,height=3.2in,angle=0]{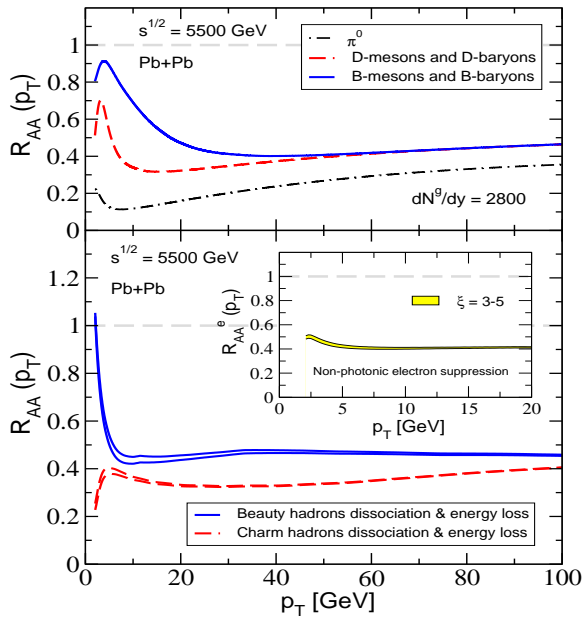}
\caption{ Suppression of $D$ and $B$ meson production in central Pb+Pb
collisions at $ \sqrt{s_{NN}} =5500$~GeV at the LHC in
two different scenarios. The top panel shows the quenching of heavy hadrons
only due to partonic energy loss. The bottom panel gives $R_{AA}$ for
$D$s  and $B$s with partonic energy loss as well as 
collisional dissociation of heavy meson. Insert shows the corresponding
attenuation of non-photonic electrons in a limited $p_T$ range.} 
\label{fig:BDSupp-LHC}
\end{figure}

Results for the suppression of open heavy flavor final states  
in central Pb+Pb collisions at $\sqrt{s_{NN}}=5500$~GeV at the LHC in a medium
of soft gluon rapidity density $dN^g/dy=2800$ are presented in Fig.~\ref{fig:BDSupp-LHC}.   
The top panel illustrates the effect of partonic energy loss with  
$R_{AA}^{B}(p_T) = R_{AA}^{D} (p_T) $ for $p_T$ about and above $50$~GeV. $D$ and $B$ 
meson quenching approaches the suppression level of light hadrons only 
at very high transverse momentum due to the significant gluon fragmentation component to
pion production in this kinematic region. The bottom panel shows the attenuation of open 
heavy  flavor when we include collisional dissociation in the medium. The most important 
feature of our results, when compared to traditional jet quenching studies, is that at 
the LHC  $R_{AA}^{B}(p_T) \approx R_{AA}^{D} (p_T) $ at a much lower $p_T \simeq 10$~GeV.
The nuclear modification factor of non-photonic electrons is also shown for completeness 
in the insert of Fig.~\ref{fig:BDSupp-LHC}. By comparing the top and bottom panels
of Fig.~\ref{fig:BDSupp-LHC}  we conclude that only experiments at the LHC 
will have the transverse momentum coverage to fully explore heavy flavor dynamics in 
the QGP. We finally remark that the small difference
in the suppression of $D$ and $B$ mesons at $p_T \sim 50$~GeV is from the interplay
of two quenching mechanisms  -  meson dissociation and parton energy loss. This 
difference disappears at $p_T \sim 150$~GeV.

%%%%%%%%%%%%%%%%%%%%%%%%%%%%%%%%%%%%%%%%%%%%%%%%%%%%%%%%%%%%%%%%%%%%%%%%%%

\vspace*{0.5cm}

\section{Conclusions}
\label{conclude}

Detailed theoretical studies have shown that partonic energy loss
of heavy quarks in the hot QGP~\cite{Wicks:2007am}  cannot explain 
the large suppression of non-photonic electrons observed by
PHENIX~\cite{Adare:2006nq}  and STAR~\cite{Abelev:2006db} collaborations   
in central Au+Au collisions at $\sqrt{s_{NN}}=200$~GeV per nucleon pair. 
This discrepancy can naturally be resolved if collisional dissociation of 
heavy mesons that tend to form  inside the QGP is taken into 
account~\cite{Adil:2006ra}. In this paper
we reported first results from an approach that attempts to combine charm and 
beauty quark quenching with $D$ and $B$ meson inelastic breakup processes 
with  the goal of describing open heavy flavor production in nucleus-nucleus 
collisions over the full $p_T$ range that will be experimentally 
accessible at RHIC and the LHC. The treatment of cold nuclear matter 
effects for massive final state mesons was brought
on par with that for pions and photons~\cite{Vitev:2008vk} and the 
Cronin enhancement at transverse momenta $\sim 4$~GeV was found to be the
most important. To put studies of heavy meson formation and dissociation
in the plasma~\cite{Adil:2006ra,Dominguez:2008be} on firmer theoretical
ground we investigated in the framework of potential models~\cite{Mocsy:2007yj}
the existence of such bound states in the vicinity of $T_c$.
We found that while the temperature at which $D$ and $B$ mesons cease to form
depends on the details of the in-medium  quark-antiquark potential and the 
light quark mass, as a rule bound states survive well above $T_c$. Using 
the light-cone description of hadrons~\cite{Brodsky:1997de} and
the operator definitions of distribution and decay probabilities from 
factorized perturbative QCD~\cite{Collins:1981uw} 
we calculated the charm and beauty PDFs and FFs in a co-moving plasma.

In the instant wavefunction approximation, relevant to an 
out-of-equilibrium jet propagation through the medium when the timescale for
the onset of thermal effects exceeds $\tau_{\rm form}$, we evaluated the 
$D$ and $B$ meson cross section suppression from partonic and hadronic interactions 
in the QGP. We  found that for $p_T > 4$~GeV at RHIC and $p_T > 10$~GeV at the LHC 
$ R_{AA}^B(p_T) \approx R_{AA}^D(p_T)$  and the transverse  momentum dependence
of this attenuation is weak. In our study meson dissociation played a dominant role 
up to  $p_T \sim 10$~GeV for open charm and up to $p_T \sim 30$~GeV for open beauty 
hadrons. To obtain a good  description of the non-photonic electron 
suppression in Au+Au collisions at RHIC after the inclusion of Cronin 
enhancement~\cite{Accardi:2002ik} approximately 50\%  larger collisional 
broadening~\cite{Vitev:2003xu} that leads to the meson breakup was required.
Predictions were given for the  $D$ and $B$ hadron and the corresponding 
non-photonic electron quenching in Cu+Cu and Pb+Pb collisions at 
RHIC and the LHC, respectively, that will very soon be confronted by 
data. We conclude by emphasizing that while charm quark/meson dynamics
in the medium is well within the reach of RHIC experiments, for a comprehensive and 
definitive study which can disentangle the medium effects on beauty quarks/mesons 
not only the enhanced cross sections but also the extended  
$p_T$ reach at LHC experiments will play a critical role.

\begin{acknowledgments}
This research is  supported by the US Department of Energy, Office
of Science, under Contract No. DE-AC52-06NA25396 and in part by the LDRD program
at LANL, the NNSF of China and the MOE of China under Project No. IRT0624.
\end{acknowledgments}

\begin{appendix}

\section{Notation and quantization of quark and gluon fields}
\label{Quantize}

To quantize the theory on the light cone we choose the light-cone time as
$x^{+}=(1/\sqrt{2})(x^0+x^3)$. The light-cone ``spatial'' components are taken
as $(x^{-},{\bf x})$ where $x^{-}=(1/\sqrt{2})(x^0-x^3)$ and ${\bf x}$ is a two
component spatial vector with $({\bf x})^{ 1}=x^1$ and $({\bf x})^{ 2}=x^2$. With this
definition, the dot product between two four vectors $a$ and $b$ has the form
$a\cdot b = a^{+}b^{-} + a^{-}b^{+} - {\bf a}\cdot {\bf b}$. We use the notation where $\vec{a}$
stands for the spatial components of a four vector $a^\mu$, and the $3$-tuple 
$\vec{a}^+ = (a^+,{\bf{a}})$.

For quarks of every
flavor and color there is a corresponding Dirac field $\psi_{cf}(x)$. In what
follows, we will suppress the color index $c$ and the flavor index $f$ for
simplicity. The free field Lagrangian for $\psi$ is given by:
\begin{equation}
{\cal{L}}=\bar{\psi}\bigl(i\dslash-m\bigr)\psi\label{psi lagrangian}\; ,
\end{equation}
where $\dslash=\gamma^\mu\partial_\mu$. $\gamma^\mu$ are Dirac matrices
satisfying $\{\gamma^\mu,\gamma^\nu\}=2\eta^{\mu\nu}$. We write the Dirac matrices
in the chiral representation:
\begin{equation}
\gamma^0=\left(\begin{array}{cc}
0 & 1\\
1 & 0 
\end{array}\right),\; 
\gamma^i=\left(\begin{array}{cc}
0 & \sigma^i\\
-\sigma^i & 0 
\end{array}\right){\rm{for}}\;i=1,2,3\label{gamma conventions}\;.
\end{equation}
The light-cone Dirac matrices are defined by the relations
$\gamma^{\pm}=(1/\sqrt{2})(\gamma^0\pm\gamma^3)$, and  ${\bm{\gamma^{
i}}}=\gamma^i$ for $i=1,2$.

The equation of motion for $\psi$ is simply
\begin{equation}
(i\dslash-m)\psi(x)=0\label{psi equation}\;.
\end{equation}
We look for plane wave solutions of form $u(p) \exp(-ip\cdot x)$ and 
$v(p) \exp(ip\cdot x)$ with $p^{+}>0$. To satisfy the equation of motion the 
four momentum $p$ should be on shell. In addition, the Dirac spinors $u(p)$ and $v(p)$
satisfy the following equations:
\begin{eqnarray}
%\begin{split}
&& \bigl(\pslash{p} - m\bigr)u(p)=0 \;, 
\quad \bigl(\pslash{p} + m\bigr)v(p)=0\label{v equation}\;.
%\end{split}
\end{eqnarray}
The solutions to these equations can be written a nice form, which we use in the calculation
of the heavy quark fragmentation functions: 
\beqar
u(p,s) = \frac{1}{\sqrt{2(p^0+m)}}
\left(
 \begin{array}{c}
( p \cdot \sigma + m ) \xi^u_s \\ 
(p \cdot \bar{\sigma} + m) \xi^u_s   \end{array} 
\right)\; ,
\eeqar{spinoru}
\beqar
v(p,s) = \frac{-\gamma^5}{\sqrt{2(p^0+m)}}
\left(
 \begin{array}{c}
( p \cdot \sigma + m ) \xi^v_s \\ 
(p \cdot \bar{\sigma} + m) \xi^v_s   \end{array} 
\right)\; ,
\eeqar{spinorv}
where $\xi_s$ is  a two dimensional unit vector, which is different for $u$ and $v$
and depends on the chosen representation. For example, based on the choice of $\xi^u_s$
and $\xi^v_s$,  $u(p,s)$ and $v(p,s)$ may or may not be helicity eigenstates.
In our calculations we only use the following properties:
$\xi^{u,v\dagger}_s \xi^{u,v}_{s'} = \delta_{ss'}$ and $\sum_{s,s'} \xi^{u}_{s}
\xi^{v\dagger}_{s'}\varepsilon_{ss'} = 
- 1_{2\times 2}$.
$\sigma^\mu = (1,\vec{\sigma})^\mu$ and $\bar{\sigma}^\mu = 
(1,-\vec{\sigma})^\mu$ in the instant form,
where $\vec{\sigma}$ are the usual Pauli matrices.

The general solution to the classical equation of motion can then be written as:
\begin{equation}
\psi(x)= \int \frac{dp^+ d^2{\bf p}}{(2\pi)^3 2p^+ } \bigl(a_{ps}u(p,s) e^{-ip\cdot
x}+b^\dagger_{ps}v(p,s) e^{ip\cdot x}\bigr)\label{psi solution}\;,
\end{equation}
where we will soon identify $a_{ps}$ and $b^\dagger_{ps}$ as creation and annihilation
operators.

In light-cone coordinates, not all the components of $\psi$ are dynamical,
meaning that the time derivative of some components do not appear in the
Lagrangian. To separate these, one defines projection operators 
$\Lambda_{+}$ and $\Lambda_{-}$ as follows.
\begin{equation}
\Lambda_{\pm}=\frac{1}{2}\gamma^{\mp}\gamma^{\pm}
=\frac{1}{\sqrt{2}}\gamma^{0}\gamma^{\pm}=\frac{1}{\sqrt{2}}\gamma^{\mp}\gamma^{0}\;.
\end{equation}
They are hermitian matrices satisfying $\Lambda_{+}+\Lambda_{-}=1$, 
$\Lambda_{\pm}^2=\Lambda_{\pm}$. One defines the ``plus'' and ``minus'' components
of $\psi$ as the projections $\psip=\Lambda_{+}\psi$ and
$\psim=\Lambda_{-}\psi$, respectively. Using the relation
$\gamma^{+}\Lambda_{-}=0$ it is easy see that $\psim$ is non-dynamical.
Therefore, one can use the equations of motion to write it in terms of the
dynamical field $\psi_{+}$, which has the following form:
\begin{equation}
\psip(x) =  \int \frac{dp^+ d^2{\bf p}}{(2\pi)^3 2p^+ }  \bigl(a_{ps}\up(p,s) e^{-ip\cdot
x}+b^\dagger_{ps}\vpp(p,s) e^{ip\cdot x}\bigr)\label{psiplus solution}\;.
\end{equation}
where $\up(p,s)=\Lambda_{+}u(p,s)$ and
$\vpp(p,s)=\Lambda_{+}v(p,s)$. They have norms given by:
\begin{equation}
\up^{\dagger}(p,s)\up(p,s')=\vpp^{\dagger}(p,s)\vpp(p,s')=\sqrt{2}p^{+}\delta_{ss'}\;.
\end{equation}

The canonical anticommutation relations for the $\psip$ and  $\psip^\dagger$ fields are:
\begin{equation}
\begin{split}
\{ \psip(x^{+},x^{-},{\bf x}),& \psip^\dagger(x^{+},y^{-},{\bf y})\}=\\
&\frac{\Lambda_{+}}{\sqrt{2}}\delta(x^{-}-y^{-})\delta^{(2)}({\bf x}-{\bf y})\;.
\end{split}
\end{equation}
The anticommutators of two fields or their hermitian conjugates vanish.
Writing $\psi$ in terms of the $a$ and $b$ operators, Eq.~(\ref{psiplus solution}), we find that,
\begin{equation}
\begin{split}
\{a_{ps},a^\dagger_{qs'}\}=\delta_{ss'} 2 p^+ (2\pi)^3 \delta(p^+-q^+)\delta^{(2)}({\bf p}-{\bf q})\;,\\
\{b_{ps},b^\dagger_{qs'}\}=\delta_{ss'} 2 p^+ (2\pi)^3 \delta(p^+-q^+)\delta^{(2)}({\bf p}-{\bf q}) \;,
\end{split}
\end{equation}
and that all other anticommutator combinations should be zero. Therefore $a$ and $a^{\dagger}$ 
can be interpreted as annihilation and creation operators for the fermions, respectively, and $b$
and $b^{\dagger}$ as annihilation and creation operators for the antifermions.

Finally, we describe the quantization of the gauge field. The free field
Lagrangian is given by: 
\begin{equation}
{\cal{L}}=-\frac{1}{4}F_{\mu\nu}F^{\mu\nu}\label{EM lagrangian}\;,
\end{equation}
where $F$ is the color field strength tensor. The main new feature
that arises during quantization is that we need to fix a gauge. We choose the
light-cone gauge, $A^{+}=0$, consistent with the gauge choice for the diagrammatic calculations. 
In this gauge the operator $A^{-}$ is non dynamical and can be eliminated using the equations 
of motion,
\begin{equation}
A^{-}=\frac{-1}{\partial_{-}} {\bm \partial} \cdot {\bf A} \label{Aminus equation}\;.
\end{equation}
The dynamical degrees of freedom are the ${\bf A}$ fields.
We define
two polarization vectors with perpendicular components $ {\bm \epsilon}(+1) =
(1/\sqrt{2})(-1,-i)$ and $ {\bm \epsilon}(-1) = (1/\sqrt{2})(1,-i)$. The quantized
operator ${\bf{A}}$ can then be written as
\begin{equation}
{\bf A}(x)=  \int \frac{dp^+ d^2{\bf p}}{(2\pi)^3 2p^+ } 
\bigl(c_{p\lambda}{\bm \epsilon}(\lambda) e^{-ip\cdot
x}+c^\dagger_{p\lambda}{\bm \epsilon}^*(\lambda) e^{ip\cdot x}\bigr)\;.
\end{equation}
with $\lambda=\pm 1$. The canonical commutation relations are,
\begin{equation}
\begin{split}
[\partial_-{\bf A}^{ i}(x^{+},x^{-},{\bf x}),& \, {\bf A}^{
j}(x^{+},y^{-},{\bf y})]=\\
&\frac{-i\delta^{i,j}}{2}\delta(x^{-}-y^{-})\delta^{(2)}({\bf x}-{\bf y})\;,
\end{split}
\end{equation}
which imply:
\begin{equation}
[c_{p\lambda},c^{\dagger}_{q\lambda'}]=\delta_{\lambda\lambda'}
2 p^+ (2\pi)^3 \delta(p^+-q^+)\delta^{(2)}({\bf p}-{\bf q}) 
\;.
\end{equation}
$c_{p\lambda}$ annihilates as gluon and $c^\dagger_{p\lambda}$ creates a gluon.
The four-component gauge field $A^\mu=[A^{+},A^{-},{\bf A}]$ can be written compactly by introducing the four
component polarization vectors $\epsilon(p,\pm 1)=[0,({\bm \epsilon}(\pm 1)\cdot
{\bf p})/p^{+},{\bm \epsilon}(\pm 1) ]$, which
satisfy $p\cdot\eps{p}{\lambda}=0$:
\begin{equation}
A^{\mu}(x)= \int \frac{dp^+ d^2{\bf p}}{(2\pi)^3 2p^+ } \bigl(c_{p\lambda}{\eps{p}{\lambda}} e^{-ip\cdot
x}+c^\dagger_{p\lambda}{\epsstar{p}{\lambda}} e^{ip\cdot x}\bigr)\;.
\end{equation}

With the creation operators for the particles in hand, we can generate the
multi-parton  Fock states. A single quark state can be written as: 
\begin{equation}
|\vec{p}^+, \alpha \rangle  = a^\dagger _{\alpha}( \vec{p}^+ ) | 0 \rangle\;,
\end{equation}
where the creation operator $a^{\dagger}$ will carry color, spin/helicity/polarization and 
quark flavor indices $\alpha$, as appropriate.
These are normalized as:
\begin{equation}
\langle \vec{q}^+, \alpha' | \vec{p}^+, \alpha \rangle 
 = \delta_{\alpha \alpha'}  2p^{+} (2\pi)^3 \delta^{(3)} ( \vec{p}^+ - \vec{ q}^+) \; .
\end{equation}

\section{Meson wavefunction in instant and light-cone forms}
\label{Instant}

In this section we describe the relation between the light-cone meson wavefunction 
and the meson wavefunction in instant form. We do this explicitly for the case of the lowest order 
Fock  component that  consists of a heavy quark $Q$ and a light antiquark $\bar{q}$. In the 
instant form, in the hadron rest frame, we write the meson wavefunction as:
\begin{equation}
\begin{split}
|&m_h,{\vec{0}}\rangle_I =
\sqrt{2m_h} \int \frac{d^3\vec{q_1} }{\sqrt{(2\pi)^3}} \frac{d^3\vec{q_2}}{\sqrt{(2\pi)^3}}
\sqrt{\frac{1}{4E_1E_2}}f({\vec{q_1}}) \\
& \times \delta^{(3)} ({\vec{q_1}}+{\vec{q_2}})
\frac{\delta_{c_1c_2}}{\sqrt{3}}
\frac{M_{s_1s_2}}{\sqrt{2}}{a}_Q^{\dagger\,c_1s_1 }({\vec{q_1}})
{b}_q^{\dagger\,c_2s_2 }({\vec{q_2}})|0\rangle\;,
\end{split}
\end{equation}
where $E$ refers to the on-shell energy $E_i=\sqrt{{\vec{q}_i}^2+m_i^2}$ and 
$E_h=\sqrt{m_h^2+{\vec p}^2}$ is simply $m_h$ in
the rest frame. Here, $f({\vec{q}})$ represents the momentum space 
wavefunction normalized such that,
\begin{equation}
\int \frac{d^3\vec{q}}{(2\pi)^3} |f({\vec{q}})|^2 = 1 \; .
\label{fqnorm}
\end{equation}

In this manuscript we concentrate on distributions $f({\vec{q}}^2)$ of spherically-symmetric type,
relevant to $^1S_0$ and $^3S_1$ open heavy flavor states. The
particular example for our case will be a harmonic oscillator distribution 
$f \sim \exp(-\vec{q}^2 a_0^2/2)$. The invariant mass squared of the multi-parton 
system of no net transverse momentum (${\bf P}=0$) is given by:
\beq
M^2(\vec{q}) = 2 \left(\sum_j p^+_i\right) \left(\sum_i p^-_i\right)  = 
\sum_i \frac{m_{\perp\, i}^2}{x_i} \; ,
\eeq{invmass}
where $m_{\perp i}^2=m_i^2+{\bf{q}}_i^2$.
%The light-cone representation is frame independent and in the rest frame where $f({\vec{q}})$ 
%is calculated:
%\beq
%x_i = \frac{(E_i+q^z_i)/\sqrt{2}}{m_h/\sqrt{2}}\;, \quad  1 = \sum_i x_i   = 
%\frac{1}{m_h}\sum_i E_i   \; .
%\eeq
We see that $M^2$ is a function of $\vec{q}$, and the mass of the hadron can be
intuitively thought of as $m_h^2=M^2(\langle\vec{q}\rangle)$. For the rest of the
section we will concentrate on the specific case of a quark-antiquark state $M^2
= {(E_1 + E_2)}^2 $. The form of $M^2$ simplifies in the following two special
cases:
%Identifying the invariant mass of the parton system with the physical mass of the 
%hadron we find:
\beqar
\label{case1}
{\vec{q}}^2 &\approx& \frac{M^2}{4} +c_1(m_1,m_2)\;, 
\quad {\rm if} \; \vec{q}^2  \gg m_i^2 \;, \\  
{\vec{q}}^2 &\approx& \frac{M^2}{\frac{(m_1+m_2)^2}{(m_1m_2)}} +c_2(m_1,m_2)\;, 
\quad {\rm if} \; \vec{q}^2  \ll m_i^2 \;. \quad
\label{case2}
\eeqar{cases}  
In all cases the same functional dependence on  ${\bf k} = {\bf q}$ and $x$ is obtained
via $\vec{q}^2 \sim M^2$  and the numerical coefficients $c_i(m_j)$ are 
absorbed in the proper normalization of 
hadron state. This yields: 
\beq
\psi \sim  \exp\left( - \frac{{\bf k}^2 + (1-x)m_1^2 + x m_2^2}{2 \Lambda^2 x(1-x) }   
  \right) \; .
\eeq{expform}
In two limiting cases the coefficient that relates $\Lambda$ in Eq.~(\ref{expform})  
to $a_0$ (the width of $f(\vec{q}^2)$) can be analytically estimated, 
as discussed above. Since the form of
$\psi$ is the same in the two extreme cases, we find it more convenient to treat $\Lambda$ 
as a constant which can be determined through the requirement that we obtain the same 
expectation value for the transverse
momentum from the instant and light-cone wavefunction forms, i.e. 
$\langle {\bf k}^2 \rangle = \langle {\bf q}^2 \rangle$. For the example of typical 
open heavy flavor 
wavefunctions~\cite{Adil:2006ra} the numerical values are within 30\% of the
values for the two limiting cases (\ref{case1}, \ref{case2}).
% More precisely, $\Lambda$ 
% can be determined through the requirement that we obtain the same $\langle {\bf k}^2 \rangle = \langle {\bf q}^2 \rangle$  
% from the instant and light-cone wavefunction forms. For the example of typical open heavy flavor 
% wavefunctions~\cite{Adil:2006ra} the numerical values are within 30\% of the analytic estimates. 

To establish the correspondence between the instant-form wavefunction and the light-cone 
wavefunction, we now convert the spatial momentum integration variables ${\vec{q}}$ 
to ${\vec{q}^+}=(q^+,{\bf q})$. The Jacobian for the transformation is $q^+/E({\vec{q}})$. 
To recover the (anti)commutation relations outlined in Appendix~\ref{Quantize} 
we find that no additional factors arise for the creation and annihilation operators. 
The proper normalization of states $\propto q^+$ is ensured when we observe that $\delta(q^z-q^{z'})
=  (q^+/E({\vec{q}}) ) \, \delta(q^+-q^{+'}) $. 
Similar relations can be established for antiquarks and gluons. One can also rewrite  
$\delta(q_1^z+q_2^z -P^z)  \rightarrow (\sqrt{2}) \, \delta(q_1^++q_2^+-P^+)$ in the boosted
frame where we will match the instant and light-cone forms of the state. 
$P^+$ is the plus component of the light-cone momentum of the meson. Note that 
in the meson rest frame $P^z=0$.  Making these variable changes and recalling that 
${\bf q} = {\bf k} $ leaves us with the same meson state: 
\begin{eqnarray}
&& | {\vec{P}}^+ \rangle  =\int\prod_{i=1}^{2} 
\frac{d^2{\bf k}_id x_i}{\sqrt{(2\pi)^3} \sqrt{2x_i}} 
\; \delta\left(\sum_i x_i - 1\right) \nonumber \\ 
&& \times \delta^{(2)}\left(\sum_i {\bf k}_i\right)  
2  {\sqrt{E_h}}  \sqrt{\frac{E_1E_2}{q^+_1q^+_2} }
\frac{M_{s_1s_2}}{\sqrt{2}} \frac{\delta_{c_1c_2}}{\sqrt{3}}\, 
\nonumber \\
&& \times f({\vec{q}}^2(x_i,{\bf k}_i)) \, {a}_Q^{\dagger\,c_1s_1 }({\vec{q_1}^+})
{b}_q^{\dagger\,c_2s_2 }({\vec{q_2}^+})|0\rangle\; .
\label{ItoLC1}
\end{eqnarray}%{chvar}
We complete the correspondence by observing:
\begin{eqnarray}
&&\lim_{E_h\gg m_h}  2  {\sqrt{E_h}}  \sqrt{\frac{E_1E_2}{k^+_1k^+_2} } \,   
f({\vec{q}}^2(x_i,{\bf k}_i)) \,  \nonumber \\
&& \qquad \quad \rightarrow \sqrt{{P^+}{\sqrt{2}} } \, f(x_i,{\bf k}_i)  \equiv 
\psi(x_i,{\bf k}_i) \; .  \label{equivalence}
\end{eqnarray}%{equivalence
The prefactor of $\sqrt{P^+}$ in front of the rest frame wavefunction $f(\vec{k})$ 
in the relation between $\psi(x_i,{\bf{k}}_i)$ and $f(\vec{k})$ 
(Eq.~\ref{equivalence}) can be physically understood as follows. 
A boost from the rest frame  to a frame  with large momentum $P^+$ 
leaves the transverse shape of the wavefunction unchanged. However, in the
large light-cone momentum direction the wavefunction is stretched in 
momentum space by a factor $\sqrt{P^+}$.
Note that Eq.~(\ref{equivalence}) combined with Eq.~(\ref{fqnorm}) yield 
$\int d^2{\bf k} dx |\psi(x,{\bf k})|^2 = 2(2\pi)^3$ as  employed in this paper.

\section{Medium contribution to heavy quark fragmentation functions}
\label{pickup}

In Sections~\ref{hqfrag} and~\ref{potmod} we calculated the fragmentation
functions for $c$ and $b$ quarks into $D$ and $B$-mesons respectively, from the
definition given in~\cite{Collins:1981uw}, Eq.~(\ref{hqFF}). In the adiabatic
picture, where the meson wavefunction has sufficient time to equilibrate with
the medium, we take the modification of $D$ and $B$ meson wavefunctions in the
medium into account by using the thermal wavefunction calculated from potential
models in Section~\ref{potmod}.

In this equilibrium situation, one can attempt to go a step further by 
estimating additional corrections that arise from thermal parton pick-up   
as follows. Instead of taking the expectation value in vacuum, we can define 
this thermal fragmentation as: 
%\begin{equation}
%\begin{split}
\begin{eqnarray}
D^T_{h/Q}(z) = z \int \frac{dy^-}{2\pi} e^{i\frac{p^+}{z}y^-}\frac{1}{3}
{\rm Tr}_{color}\frac{1}{2}{\rm Tr}_{Dirac} \frac{\gamma^+}{2}\;\; \qquad \;\;\label{hqFFtherm}
\\ \nonumber
\times\frac{\sum_m \exp(-\beta E_m) \langle m |\psi(y^-,{\bf 0})
a_h^{\dagger}(P^+)a_h(P^+)\bar{\psi}(0,{\bf 0}) |m \rangle}{\sum_m \exp(-\beta
E_m)} \,. 
\end{eqnarray}
%\end{split}
%\end{equation}
We take Eq.~(\ref{hqFFtherm}) to be the minimal generalization of the definition
of fragmentation in perturbative QCD, with $z<1$.
In Eq.~(\ref{hqFFtherm}), $E_m$ is the energy of the state $|m\rangle$, $\beta$ is the
inverse temperature, and $\exp(-\beta E_m)$ gives the thermal weight of the
state $m$. For convenience of notation, we are writing the spectrum of states as
being discrete, with the states normalized as $\langle m | n \rangle = \delta_{mn}$. 
One can immediately see that in the case that the operator does
not contract with the state $m$, we get back the definition in vacuum because the sum
of exponentials cancels out between the numerator and the denominator. For
weakly interacting quasi-particles, $E_m$ is the sum of the energies of the
quasi-particles in the state $m$. In this approximation, the denominator 
\begin{equation}
\begin{split}
Z = \sum_m \exp(-\beta E_m)
 = \Pi_{\varepsilon_i} \bigl(\exp(\beta\varepsilon_i)\pm1\bigr)\label{Z}\;,
\end{split}  
\end{equation}
where the $+$ is for Fermions and $-$ for the bosons. When the operator 
contracts with one quasiparticle with energy
$\varepsilon$ from the medium (equivalent to saying that the meson picks up a
thermal parton of energy $\varepsilon$), all the other terms in the product
(Eq.~\ref{Z}) cancel out as before, except the one corresponding to energy
$\varepsilon$. Hence, one gets an additional factor of
$g(\varepsilon)=1/\bigl(\exp(\beta\varepsilon)\pm1\bigr)$ depending on whether the parton
is fermion $(+)$ or a boson $(-)$.

\begin{figure}[!t]
\vspace*{.2in}
\includegraphics[width=3.0in,height=1.4in,angle=0]{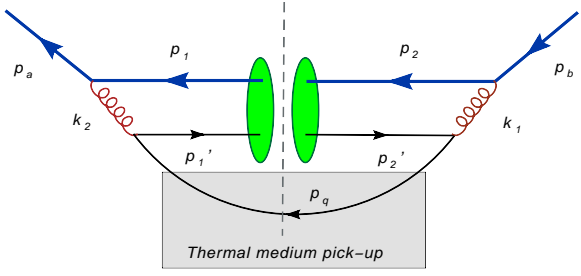} \\
%\vspace*{-1.in}
\caption{Thermal correction to the fragmentation function that arises from an in-medium
quark pick-up.}
\label{pQCDFFtherm}
\end{figure}

At the same order in $\alpha_s$ as vacuum fragmentation the following Feynman
diagram is the most important additional correction to $D_{h/Q}(z)$
function, Fig.~\ref{pQCDFFtherm}. We call this correction ${{D}_{h/Q}}^{(1)}(z)$. 
This diagram can be interpreted as follows. The heavy quark 
picks up a light thermal antiquark from the medium and forms a meson. The evaluation of
this diagram is very similar to Eq.~(\ref{eq:hqFF-2}) and yields:
\begin{eqnarray}
& & \!\!\!D^{(1)}_{h/Q}(z) =\int \frac{dx_1 d^2{\bf k}_1 \psi(x_1, {\bf k}_1) }
{(2\pi)^32\sqrt{x_1(1-x_1)} }
\frac{dx_2 d^2{\bf k}_2 \psi^*(x_2, {\bf k}_2) }
{(2\pi)^32\sqrt{x_2(1-x_2)} } 
\nonumber \\
& & \!\!\!  \frac{M(j)_{s_1 s_1^{\prime}} }{\sqrt{2}}
\frac{M(j)_{s_2 s_2^{\prime}} }{\sqrt{2}}
\int ds \; \theta \left(s-\frac{m_h^2}{z}-\frac{m_q^2}{1-z} \right)
\frac{\alpha_s^2 C_F^2}{3}   \nonumber \\
& & \!\!\! {\rm Tr} \Big[ \gamma^+ \frac{i}{\gamma \cdot p_a   - m_Q}
\gamma^{\mu} u_{s_1}(p_1) \bar{v}_{s_1^{\prime}} (p_1^{\prime})
\gamma^{\nu} ( \gamma \cdot p_q + m_q)
\gamma^{\sigma} \nonumber \\
& & \!\!\! v_{s_2^{\prime}}(p_2^{\prime}) \bar{u}_{s_2}(p_2) \gamma^{\lambda}
\frac{i}{\gamma \cdot p_b - m_q} \Pi_{\mu\nu}(p_a-p_q)\Pi_{\sigma
\lambda}(p_b-p_q)\nonumber\\
& &g(p_q^0)\Big]\frac{1}{{\rm Tr}[\gamma^+ (\gamma \cdot p) ] } \, , 
\label{eq:hqFF-2therm}
\end{eqnarray}
where $p_a = p_b = p_Q$ and $p_q$ is the momentum of the light antiquark that is picked 
up from the medium. $g(p_q^0)=\frac{1}{e^{\beta p_q^0}+1}$ is the thermal probability to 
find a quark with energy $p_q^0$ in the medium.
For future use we define the hadron momentum $p_h=p_1+p_1'=p_2+p_2'$.
The expression, Eq.~(\ref{eq:hqFF-2therm}), differs from $D_{h/Q}(z)$ only 
in the extra factor $g(p_q^0)$. To get a rough estimate of the correction, 
we replace $p_q^0$ by its typical value $\bar{p}_q^0$ and obtain,
\begin{equation}
\begin{split}
{{D}}^{(1)}(z) \sim {{D}}(z)g(\bar{p}_{{q}}^0)\;.
\end{split}
\end{equation}
>From the Feynman diagram Fig.~(\ref{pQCDFFtherm}) we  see that
$p_{{q}}=p_a - p_h$. For a typical sample of the kinematics, we consider the
following. By choosing an appropriate direction of the axes, the transverse momentum
of the hadron can be taken to be zero. Then,
\begin{equation}
\begin{split}
p_a &= \Bigl[\frac{P^+}{z}, p_a^-,{\bf{p}}_q\Bigr] \\
p_h &= \Bigl[P^+, \frac{m_h^2}{2zp^+}, {\bf{0}}\Bigr]\\
p_{\bar{q}} & = \Bigl[(1/z-1)P^+, \frac{{\bf{p}}_q^2+m_q^2}{2P^+(1/z-1)},{\bf{p}}_q\Bigr]\;.
\end{split}
\end{equation}
where ${\bf{p}}_q$ is the transverse momentum of the light antiquark and is related
to the off-shellness $(p_a^2-m_Q^2)$ of the initial heavy quark. We obtain straightforwardly: 
\begin{equation}
\begin{split}
\bar{p}_{{q}}^0&=\frac{1}{\sqrt{2}}\Bigl[P^+\bigl(\frac{1}{z}-1\bigr)+\frac{m_q^2}{2P^+(1/z-1)}\Bigr]\\
& \geq \frac{1}{\sqrt{2}}P^+(1/z-1) \; ,
\end{split}
\end{equation}
where $m_q$ is the light quark mass in the thermal medium. To obtain a
conservative estimate for the correction, we take $m_q$ to be zero.
Using typical value of $z$ for $D$ and $B$-mesons, we find that for a $4$~GeV 
$D$-meson moving through the medium at a temperature $250$~MeV, this gives $g(\bar{p}_q^0)$ 
less than $3\%$, while for a $4$~GeV $B$-meson moving through
the same medium we find corrections less than $15\%$.

\section{Analytic model of heavy quark and heavy meson multiplicities}
\label{DiffEq}

It is useful to find an approximate analytic solution to the system of ordinary differential 
equations, Eqs.~(\ref{rateq1}) and ~(\ref{rateq2}), for the purpose of understanding the 
interplay between heavy quark fragmentation and meson dissociation and evaluating the fraction 
of the time during which the charm is in a partonic state. This will help us determine the quenched
initial conditions for our numerical results. We will take the ${p_T}$-dependent
quark and hadron multiplicities, $f^Q(p_T,t) = A_Q/p_T^{n_Q} $ and $f^H(p_T,t)= A_H/p_T^{n_H}$ 
to be of power-law type~\cite{Vitev:2008vk}. We can then express: 
$$\int_0^1 dx \,  \frac{1}{x^2} \phi_{Q/H}(x) f^{H}({p}_{T}/x)  =  f^{H}({p}_{T}/c_H)\;,$$ and 
$$\int_0^1 dz \,  \frac{1}{z^2} D_{H/Q}(z) f^{Q}({p}_{T}/z) = f^{Q}({p}_{T}/c_Q) \;.$$  For any 
finite $p_T$ range the coefficients $n_H, \;n_Q, \; c_H,\; c_Q$ can be determined numerically.
In order to obtain a simple analytic solution we must also average the formation and 
dissociation rates in the interval $(0,L_{QGP})$. This yields:  
\begin{eqnarray}
\label{rateq1simp}
\partial_t f^{Q}({p}_{T},t) &=&
-  \frac{f^{Q}({p}_{T},t)}{\langle \tau_{\rm form}(p_T) \rangle} 
 +  \frac{c_H^{n_H} \, f^{H}({p}_{T},t)  }{\langle
\tau_{\rm diss}(p_T) \rangle } \;, \qquad \\[1ex]
\partial_t f^{H}({p}_{T},t) &=&
-  \frac{f^{H}({p}_{T},t) }{\langle \tau_{\rm diss}(p_T) \rangle } 
+ \frac{c_Q^{n_Q}\, f^{Q}({p}_{T},t)}{\langle \tau_{\rm form}(p_T) \rangle}
\;. \qquad
\label{rateq2simp}
\end{eqnarray}
A general solution ansatz $\propto \sum A_i e^{r_i t}$  leads to:
\begin{eqnarray}
&&r_{1,2}= \frac{1}{2}\Biggl[ { -\Bigl(\frac{1}{\langle \tau_{\rm form} \rangle} +  \frac{1}{\langle 
\tau_{\rm diss} \rangle } \Bigr) }  \\
&&  \pm \sqrt{  \Bigl(\frac{1}{\langle \tau_{\rm form} \rangle} +  \frac{1}{\langle 
\tau_{\rm diss} \rangle } \Bigr)^2 
-\frac{4}{ \tau_{\rm form}\tau_{\rm diss}  }(1- c_Q^{n_Q} {c_H^{n_H})  }  } \Biggr] 
\nonumber  \;.
\end{eqnarray}{rvalues}
From the initial conditions, Eq. (\ref{eq:initial-condition}), and Eqs.~(\ref{rateq1simp}) 
and (\ref{rateq2simp}) we can also determine the time derivatives of the heavy quark and hadron
distributions at $t=0$. Solving for the coefficients in the linear superposition of independent
solutions we find: 
\begin{equation}
\begin{split}
f^{H}({p}_{T},t) =& \Biggl[ \frac{c_Q^{n_Q} \, e^{r_1 t} }{\langle \tau_{\rm form}(p_T)\rangle(r_1-r_2)} 
 \\ 
& 
-\frac{c_Q^{n_Q}\,  e^{r_2 t} }{\langle \tau_{\rm form}(p_T) \rangle (r_1-r_2)} \Biggr] f^{Q}({p}_{T},0)  
\;,~\label{solH}
\end{split}
\end{equation}
and, 
\begin{equation}
\begin{split}
f^{Q}({p}_{T},t) =& 
- \Biggl[ \frac{ e^{r_1 t} }{r_1-r_2} \Bigl( \frac{1}{\langle \tau_{\rm form}(p_T) \rangle}   +
  r_2\Bigr)
\\
& 
- \frac{ e^{r_2 t} }{r_1-r_2}  \Bigl( \frac{1}{\langle \tau_{\rm form}(p_T) \rangle}   + r_1\Bigr)
  \Biggr] \,
f^{Q}({p}_{T},0) \;.~\label{solQ}
\end{split}
\end{equation}
We can use Eqs.~(\ref{solH}) and~(\ref{solQ}) not only to study the mix of heavy quarks and hadrons as a function
of  $  \langle \tau_{\rm form}(p_T) \rangle \,, \; \langle \tau_{\rm diss}(p_T) \rangle $ but to also evaluate the
fraction of the time the charm or bottom are in a partonic state:
\begin{equation}
\frac{t_{\rm partonic}(p_T)}{t_{\rm total}} = 
\frac{ \int_0^{L_{QGP}} dt\, t\, f^Q(p_T,t)  }{  \int_0^{L_{QGP}} dt \, t \, f^Q(p_T,t)  
+  \int_0^{L_{QGP}} dt \, t\, f^H(p_T,t)   } \;.\label{fractional time}
\end{equation}

\end{appendix}

\end{document}